\documentclass[preprint]{aastex}
\usepackage{graphicx}

\received{}
\accepted{}
\journalid{}{}
\articleid{}{}
\slugcomment{}

\renewcommand{\AA}{A}
\newcommand{\etal}{{\em et~al.}}

\newcommand{\g}{{\it g}}
\renewcommand{\r}{{\it r}}
\renewcommand{\i}{{\it i}}

\newcommand{\ionsign}[1]{{\,\footnotesize\texttt{#1}}}
\newcommand{\OII}{\textrm{[O\ionsign{II}]}}
\newcommand{\OIII}{\textrm{[O\ionsign{III}]}}
\newcommand{\NII}{\textrm{[N\ionsign{II}]}}

\newcommand{\OIIdoublet}{\textrm{[O\ionsign{II}]}\,$\lambda\lambda$3727,3730}
\newcommand{\OIIIdoublet}{\textrm{[O\ionsign{III}]}\,$\lambda\lambda$4960,5008}
\newcommand{\NIIdoublet}{\textrm{[N\ionsign{II}]}\,$\lambda\lambda$6550,6585}

\newcommand{\OIIl}{\textrm{[O\ionsign{II}]}\,$\lambda$3727}
\newcommand{\OIIu}{\textrm{[O\ionsign{II}]}\,$\lambda$3730}
\newcommand{\OIIIl}{\textrm{[O\ionsign{III}]}\,$\lambda$4960}
\newcommand{\OIIIu}{\textrm{[O\ionsign{III}]}\,$\lambda$5008}
\newcommand{\NIIl}{\textrm{[N\ionsign{II}]}\,$\lambda$6550}
\newcommand{\NIIu}{\textrm{[N\ionsign{II}]}\,$\lambda$6585}
\newcommand{\SIIl}{\textrm{[S\ionsign{II}]}\,$\lambda$6718}
\newcommand{\SIIu}{\textrm{[S\ionsign{II}]}\,$\lambda$6733}
\newcommand{\Halpha}{H$\alpha$\,$\lambda$6565}
\newcommand{\Hbeta}{H$\beta$\,$\lambda$4863}
\newcommand{\Hgamma}{H$\gamma$\,$\lambda$4342}
\newcommand{\Ha}{H$\alpha$}
\newcommand{\Hb}{H$\beta$}
\newcommand{\eclass}{$\texttt{eclass}$}
\newcommand{\ecoeff}[1]{$\texttt{ecoeff}_{#1}$}

\newcommand{\ybar}{$\overline \mathbf y$}

\newcommand{\mycomment}[1]{}

\shortauthors{Gy\H{o}ry et~al.}
\shorttitle{Correlation of emission lines and continuum}

\begin{document}

\title{Correlations between nebular emission
and the continuum spectral shape in 
SDSS galaxies}

\author{
Zsuzsanna Gy\H{o}ry\altaffilmark{\ref{Eotvos},\ref{MPIA}},
Alexander S. Szalay\altaffilmark{\ref{JHU},\ref{Eotvos}},
Tam\'as  Budav\'ari\altaffilmark{\ref{JHU}}, 
Istv\'an Csabai\altaffilmark{\ref{Eotvos},\ref{JHU}},
St{\'e}phane Charlot\altaffilmark{\ref{MPA},\ref{IAP}}
}

\email{gyory@complex.elte.hu}

\altaffiltext{1}{Department of Physics, E{\"o}tv{\"o}s University,
Budapest, Pf.\ 32, H-1518, Hungary 
\label{Eotvos}}

\altaffiltext{2}{Max Planck Institute for Astronomy, K{\"o}nigstuhl 17,
69117 Heidelberg, Germany
\label{MPIA}}

\altaffiltext{3}{Department of Physics and Astronomy, The Johns Hopkins University, 3701 San Martin Drive, Baltimore, MD~21218
\label{JHU}}

\altaffiltext{4}{Max-Planck-Institute f\"ur Astrophysik,
Karl-Schwarzschild-Strasse 1, 85748 Garching, Germany
\label{MPA}}

\altaffiltext{5}{Institut d'Astrophysique de Paris, CNRS, 98 bis
boulevard Arago, F-75014 Paris, France
\label{IAP}}

\begin{abstract}

We present a statistical study of correlations and dimensionality of
emission lines carried out on a sample of over 40,000 SDSS galaxies.
 Using principal
component analysis we found that the equivalent widths of the 11
strongest lines can be well represented using three parameters. We
also explore correlations of emission pattern with the eigenspace
representation of the continuum spectrum. The observed relations are
 used to provide an empirical prescription for expectation values
and variances of emission line strengths as a function of  spectral
shape. We show that this estimation of emission lines has a sufficient 
accuracy to make it suitable for photometric applications.
The method has already
 proved useful in photometric redshift estimation 

\end{abstract}


\section{Introduction} 
\label{sec:intro}

Galaxy spectral models play an essential role in the interpretation of
observational data. It is important to be able to characterize  models with
few parameters as accurately as possible especially when working
with photometric measurements.  Our motivation for exploring the dimensionality
of the emission pattern and its correlation with the continuum spectrum
originates in photometric redshift estimation. There are basically two
different approaches  to determining redshifts of galaxies from their
multi-band photometric data. One is empirical, for example via fitting the
color-redshift relation (\citet{connolly95}), nearest neighbors
(\citet{budavari01}), Kd-tree (\citet{csabai03}) or artificial neural networks
(\citet{firth03}).  The other approach is based on template spectra. The
templates might be either empirical, like \citet{cww}, or synthetic, e.g.,
\citet{bc03} and \citet{pegase}.  The templates are redshifted and convolved with
the filter curves of the survey.  The simulated fluxes serve as a reference set
to match with the photometry of the real objects. The best match gives an
estimate of the redshift and spectral type of a galaxy.  (For an overview, see,
e.g., \citet{ csabai03}.) In many redshift studies synthetic spectra are used,
for example in COMBO-17 (\citet{wolf04}), Hyperz (\citet{hyperz}), and EAZY 
(\citet{eazy}).  Modern surveys with good signal-to-noise ratio (S/N) and resolution, such as the
co-added stripe 82 of the Sloan Digital Sky Survey (SDSS), PanSTARRS (\citet{kaiser05}), or LSST (\citet{tyson02})
make even a more general photometric parameter inversion realistic
\citet{budavari09}. Beyond redshift, one may address other physical parameters
as well, such as star formation history, metallicity, or dust.  Then an ensemble
of model spectra parameterized by a set of such observables is the best starting
point for calibrating the photometric inversion procedure.  Not all quantities of
interest will be unambiguously visible from the photometric data, but their
degeneracies, correlations and errors can be well assessed by the calibration.
Knowing the propagation of the uncertainties allows one to minimize the error
of a particular observable and thus optimize the entire method.  This concept
assumes realistic model templates.  The templates should account for all
features contributing to the integrated fluxes, i.e.,  the spectral continua as
well as the strongest emission lines.

A suitable spectral model should include the radiation of stars,
ionized gas, and the effect of dust. There are works by
\citet{stasinska96, moy01, charlot01, panuzzo03} that couple these
components. Stellar continua are usually modeled using population
synthesis models, e.g. \citet{bc03}, PEGASE 
\citep{pegase}. Emission lines in star forming (SF) H\ionsign{II} regions
are generated by photoionization codes, e.g. PHOTO \citep{photo},
CLOUDY \citep{cloudy}. In general, a particular model is defined by age,
metallicity, star formation history and initial mass function of the
underlying stellar population. Furthermore chemical composition, density and geometry of the
ionized gas as well as dust content and certain characteristics of dust also define a model. In order to
reduce the number of free parameters one usually applies simplifying assumptions 
and self-consistency constraints, i.e. empirical relations between the
physical quantities. This enables one to produce
models described with a reasonable accuracy with about three stellar and
three gas parameters.

Principal component analysis (PCA) has
proved to be a powerful tool in exploring the correlations between
emission line properties and 
continuum spectral characteristics. \citet{sodre99} 
and \citet{stasinska01} used it for statistical analysis of 
spectral features of nearby spiral 
galaxies. They identified the trends of emission line equivalent widths (EWs) as
a function of spectral type obtained by PCA. 

SDSS provides data suitable for 
statistical analyses of nebular emission on a large sample of
galaxies. There have been numerous studies addressing the physical
properties of SDSS emission line galaxies (\citet{brinchmann04},
  \citet{kauffmann03},\citet{tremonti04}).
The aim of our present study is to elucidate the statistical
description of galaxy emission lines. This is complementary to the
previous studies as it allows one to explore, for example, the
contamination of photometric magnitudes by emission lines in an
efficient way.

In this paper we perform PCA on emission line EWs of SDSS 
galaxies in order to find the minimal number of
independent parameters describing the emission line pattern with a
reasonable accuracy. We then explore their correlations with the
continuum spectrum PCA parameters and determine the most probable
emission line pattern and its variations as a function of the
emission-free continuum spectrum. These relations enable us to add
emission lines to population synthesis model spectra in an empirical
way. This
prescription was used to improve the empirical spectral 
templates used in SDSS photometric redshift estimation. 

\section{Data}
\label{sec:data}

\subsection{Description of the SDSS sample}

We studied the emission line data of galaxies selected
from the SDSS DR6  database \citep{dr6}.
The SDSS spectroscopic catalog contains galaxies of all types brighter than $17.77$ 
mag in the $r$ band \citep{strauss02} and a roughly volume-limited 
sample of luminous red galaxies with redshifts ranging up to $z\approx 0.45$ 
\citep{eisenstein01}. The spectra are taken using 3 arcsec diameter fibers, thus
in the lowest redshift galaxies the spectroscopy 
only samples the central region. 
The data include redshifts, 
spectral type, measured  characteristics of spectral lines etc.
For a technical overview of SDSS see \citet{york00}.

The spectral line characteristics published in the spectroscopic catalog 
are evaluated by an automated pipeline. 
The pipeline is able to identify 48 spectral lines.
In order to determine the fluxes the lines are fitted by Gaussian profiles. 
 The database lists the fit parameters including
position, height, width, EW and
spectral continuum flux to each line. We use these values for our analysis.

The SDSS spectroscopic catalog also provides quantities that carry
useful information on the spectral shape of galaxies in a compact
form. \citet{connolly95} showed that the spectra of galaxies form a
low-dimensional manifold. Using only a few parameters (say three) the
spectra can be described with very good precision (99\% accuracy) and
an objective spectral classification of galaxies is also possible. The
spectroscopic pipeline applies this dimensional reduction technique as
detailed in \citet{connolly99} and \citet{yip04} to obtain the
parameters, $\texttt{ecoeff}_i$, $i=0...4$ for each galaxy. They are
the weights of the first five PCA eigenspectra of the SDSS galaxies.
  The derived
quantity $\texttt{eclass}=atan(-\texttt{ecoeff}_1/\texttt{ecoeff}_0)$ 
characterizes the shape of the spectral
continuum very well. Its increasing value corresponds to the rising blue
end of the spectrum and decreasing {4000 {\AA}} break, i.e. small/large
{\eclass} values indicate early/late types. (For illustration, 
see the left panel of Figure~\ref{fig:type-emi}, which shows the
correlation of {\eclass} with the color $u-r$.)

\subsection{Sample selection}

The sample for the present study was  selected as follows. We selected
objects that were classified by the SDSS 
photometric pipeline as galaxies. We restricted
our investigations to the strongest 11 emission lines listed in
Table~\ref{tab:linenames}. In order to include only reliable data we
required the median S/N in the $g^*$ band to be more than 10,
$\texttt{sn}_0 > 10$ in the SpecObj data table of the SkyServer%
\footnote{http://skyserver.sdss.org} catalog science archive.
For all galaxies included in our dataset we
required that all lines listed in Table~\ref{tab:linenames} are
measured (125,832 objects).
In order to exclude measurements with extremely large errors
we set an upper cut in the errors of equivalent widths at 5~\AA, this
affected  5\% of the parent sample. 
Our typical errors are 
between  0.2 and 0.4\AA, depending on line.
The cleanness of data manifests e.g. in correct ratios of the known doublet lines.  
In some cases
the line profile fit does
not resolve the broad {\Halpha} and its two close neighbors {\NIIl}
and {\NIIu} correctly.
In order to remove this systematic error, we excluded objects with
blended {\Halpha}.  
An undesired effect of this sampling is the
exclusion of galaxies that have some broad component around {\Ha},
e.g. Seyfert I galaxies.
Our criterion to have all 11 lines reliably measured raises a concern     
about completeness in very low and very high metallicity ranges. In the former
group {\NII} lines might be unmeasurable, whereas in the latter group {\OIII}
might be very weak or absent. We address this question later in Section~\ref{sec:eigenspace}.
We also required $\texttt{sigma}<2$  for the {\OIIdoublet}, in order to exclude cases
where the fit did not resolve the two lines but captured the blended
doublet instead.
 We found
these particular constraints the most suitable for obtaining both 
a clean and representative dataset.

\begin{figure}
\begin{center}
\includegraphics[width=16cm]{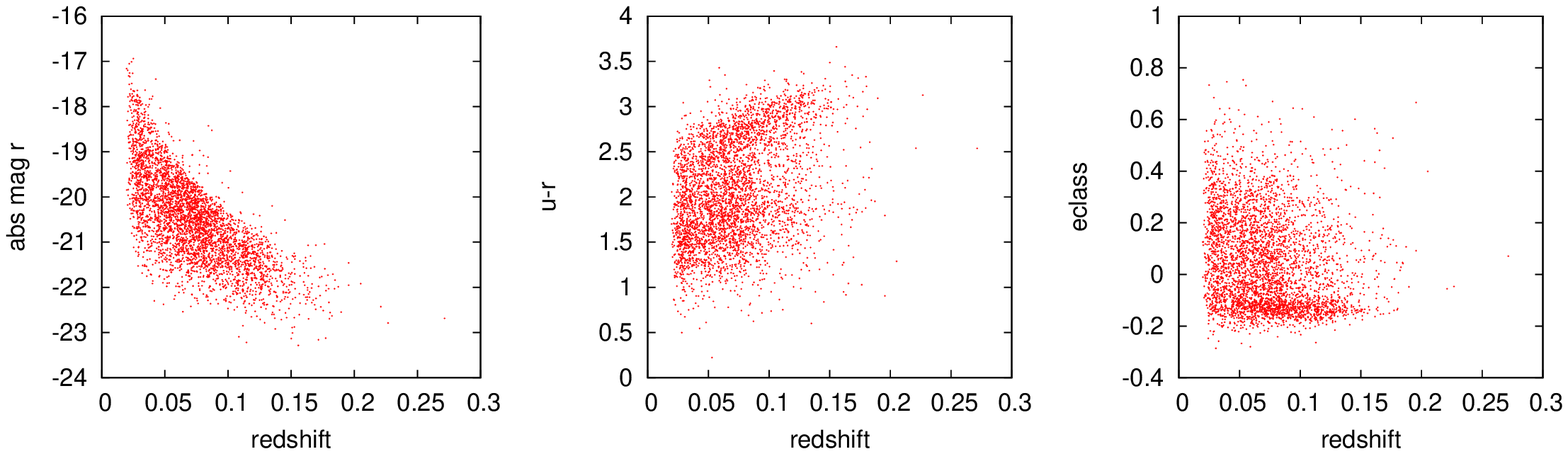}
\end{center}
\caption[]{Distribution of
absolute magnitude (left),
  color $u-r$ (middle) and the continuum spectral type parameter
  {\eclass} (right) vs. redshift  shown for a 10\% random subset  of our
  sample. 
Two distinct groups of galaxies can
  be identified in both  color and spectral  type.}
\label{fig:data}
\end{figure}

Our final sample contains 40,312 galaxies. Since our selection criteria
did not include any constraints on the sign of the emission line
fluxes, our sample includes both emission line galaxies and
objects with mainly absorption features. 
Some characteristics of the
sample are shown in Figure~\ref{fig:data}, namely distribution of the data
in redshift, absolute magnitude, color and {\eclass}. Redshift ranges up
to $z=0.3$, with an average value of $z=0.07$. In both color  $u-r$ and
continuum type {\eclass} there are two groups of objects visible: red,
early types at higher $u-r$ and low {\eclass} and 
blue, late types at smaller $u-r$
and larger {\eclass}. As described by \cite{strateva01} based on a
study of SDSS galaxies, the two
underlying groups of the bimodal distribution in color
space can be separated by a single cut at $u-r=2.22$. According to this criterion,
 41\% of our galaxies are red  and 59\% are blue. The 
corresponding distribution of spectral types is manifested in {\eclass}
too. For more details on the distributions of galaxy spectral types 
in SDSS see \cite{yip04}. We can distinguish between SF
galaxies and active galactic nuclei (AGNs)  using emission line diagnostics
 based on the line ratios  
$N2=\log (\textrm{\NIIu  /\Ha})$ and $O3=\log (\textrm{\OIIIu
  /\Hb})$, which was
introduced by \cite{BPT} (BPT). (For the distribution of our data
 on the BPT diagram see 
Figure~\ref{fig:seagull}, however, the explanation of the different symbols can be found
later in Section \ref{sec:eigenspace}.)
In order to identify the two groups in our sample we 
adopt the AGN/SF separator of \cite{kauffmann03}
\begin{equation}
O3=0.61/(N2-0.05)+1.3.
\label{eq:agn-sep}
\end{equation}
 Objects with $O3$ values above this line are classified as
AGNs, the rest are clessified as SFs. 
Based on this cut,
 nearly 50\% of the sample are SF galaxies, 18\% have  an
AGN-like emission pattern and over 32\% of the objects cannot be
 classified as either
because they have at least one of the four
lines of Equation~(\ref{eq:agn-sep}), in most cases {\Hb} having non
positive EW's. These are objects 
with weak overall emission.

\section{Analysis of spectral lines}
\label{sec:analysis}

\subsection{Equivalent width and spectral type distribution}

\begin{figure}
\begin{center}
\includegraphics[height=16cm]{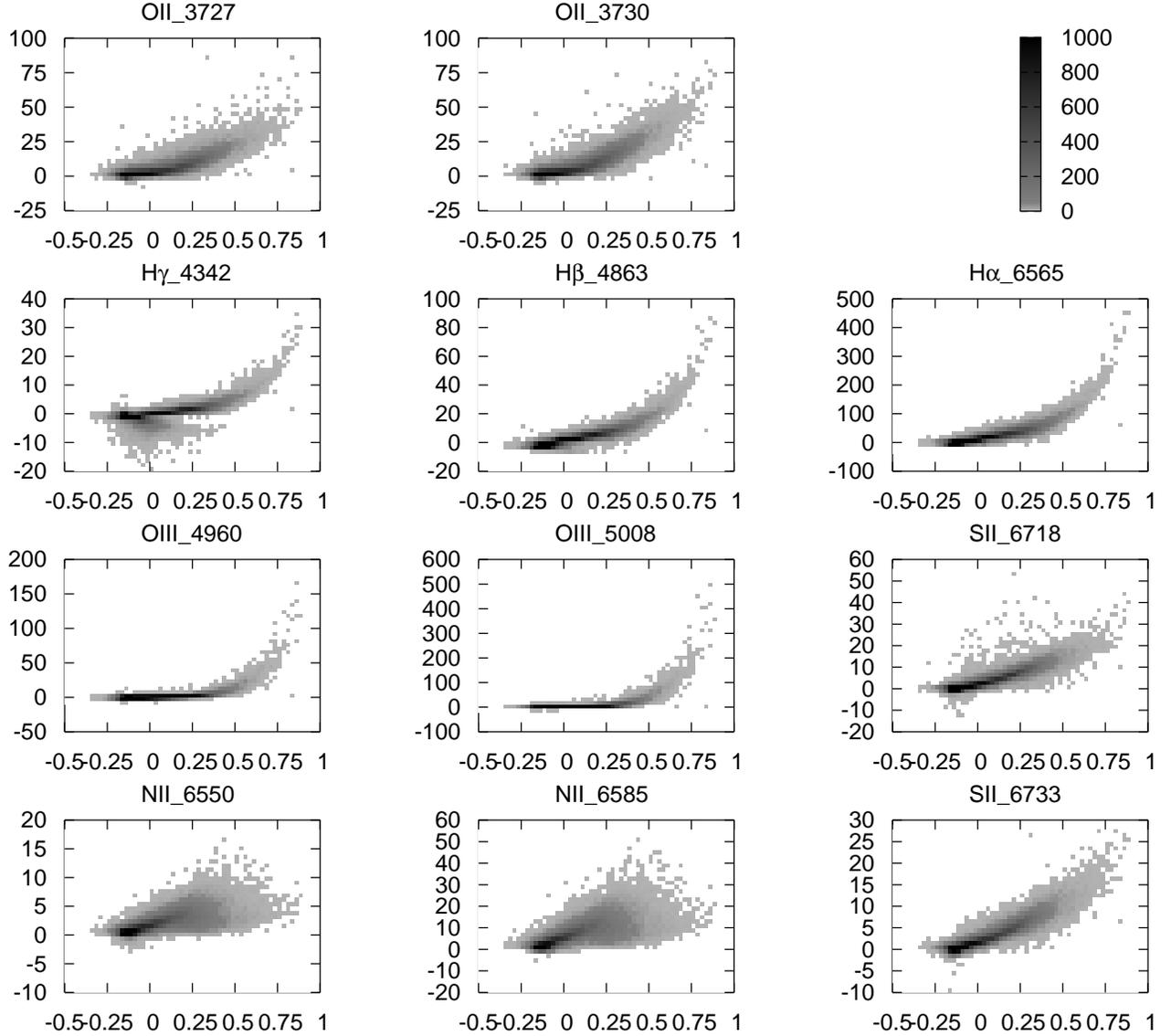}
\end{center}
\caption[]{EW -- $\texttt{eclass}$ distribution of the selected 11
  spectral lines. Galaxy counts are plotted in grayscale. The spectral type parameter
 $\texttt{eclass}$ ($x$ axis) is small/large for early/late type galaxies. 
The EWs (in \AA, $y$ axis) of all lines show a strong type dependence.
The absorption dominated early type objects are situated at negative
$\texttt{eclass}$ values. At positive $\texttt{eclass}$ the 
EW's of the emission line galaxies increase with type.}
\label{fig:ew}
\end{figure}

\begin{figure}
\begin{center}
\includegraphics[width=16cm]{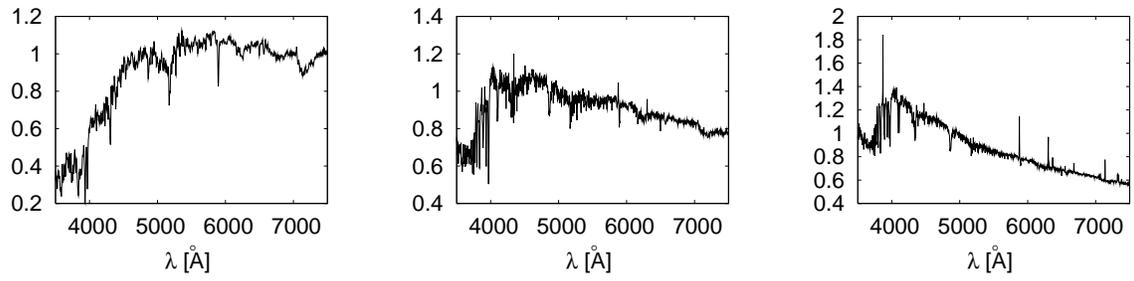}
\caption[]{Composite galaxy spectra from early type with no emission
 (left) to emission rich late type (right). }
\end{center}
\label{fig:spec}
\end{figure}

Figure~\ref{fig:ew} shows the distribution of the EW
values of our sample plotted against the type parameter
{\eclass} (the smaller the redder). The data show the
well-known 
tendency of emission lines becoming stronger 
from early to late types, see  Figure~\ref{fig:spec} for a few
examples. Star formation enhances both the blue color of late type 
galaxies and the strength of the emission lines. (This also 
makes it possible to use certain lines as measures of star formation, see e.g.
\citet{kennicutt98, kewley04}.)
This trend is obvious for all 11 analyzed lines. Early type
galaxies located at negative {\eclass} values show mainly
absorption features and almost no emission independently of
{\eclass}. This group can be distinguished from late
type galaxies situated mostly at positive {\eclass} values,
whose emission tends to rise with {\eclass}. However,
this tendency is not the same for the different lines, which
originates in different physics of formation. 
We also see galaxies that have seemingly strong absorption at
{\Hgamma} but not at {\Halpha}, which is puzzling as typically one
expects the absorption EW at {\Halpha} to be about 60\% of that at
{\Hgamma}. Since our EW's are the sum of emission { and}
absorption, this probably means that the {\Halpha} absorption is
filled up by the emission of these galaxies. This effect is 
enhanced by measurement uncertainties as well. Seeming strong absorption
values at 
 {[O\ionsign{III}]} and {[S\ionsign{II}]} are
also present due to measurement errors.

\subsection{Orthogonal approach}
\label{sec:pca}

We try to quantify the common trends and differences in the
variation of the nebular emission pattern using PCA
of the EW data. They characterize the emission 
strength physically as they do not depend on distance and the effect
of the galactic reddening is canceled by normalization. At the same
time, EW's are
affected by the intrinsic reddening caused by an inhomogeneous
distribution of star, gas and dust components \citep{calzetti94,
  charlot00}. Because we are interested in data as they are observed in
photometric measurements, we
choose not to correct for the intrinsic reddening. 

PCA is a linear transformation of data vectors
to the eigensystem of their correlation matrix. It results in an
uncorrelated representation of the data, and makes it possible to identify
the most relevant directions of variation by ranking them according to their 
information content. 

PCA of our data was carried out as follows. We represented the
EW's of each galaxy with an $M=11$ dimensional vector $\mathbf y$, with the
average {\ybar} subtracted. Diagonalizing the covariance matrix we
obtained the orthonormal set of $M$ eigenvectors or principal
components (PCs) $\mathbf e^k$. We ordered them by their eigenvalues
$\lambda^k$ (normalized to unit trace)  since they express the relative
information content of each eigenvector.  For each galaxy the
expansion coefficients of the vector $\mathbf y$ form the new $M$
dimensional vector $\mathbf c$:
\begin{equation}
\mathbf{y} = \sum_{k=1}^{M} c_k \, \mathbf{e}^k.
\label{eq:exp}
\end{equation}
The transformation of vector $\mathbf y$ to vector $\mathbf c$ corresponds 
to a simple rotation of the data vectors to the basis where their correlation matrix is diagonal.
 Inverting the transformation we obtain the original vectors again. However,
if we truncate the expansion coefficients at some $m<M$, 
the data will not be exactly restored.
The effect of omitting the $k$th principal component is the reduction
of the variance of the truncated EW estimator 
\begin{equation}
 \mathbf{y}
^{(m)} = \sum_{k=1}^{m} c_k \, \mathbf{e}^k,
\label{eq:rec}
\end{equation}
by $\lambda^k$ with respect to the original distribution of the data. 
Hence, the eigenvalues are actually a measure of the importance of each eigenvector
to reconstructing the real distribution of the data.

\subsection{The principal components}
\label{sec:eigenvectors}

\begin{figure}
\begin{center}
\includegraphics[width=12cm]{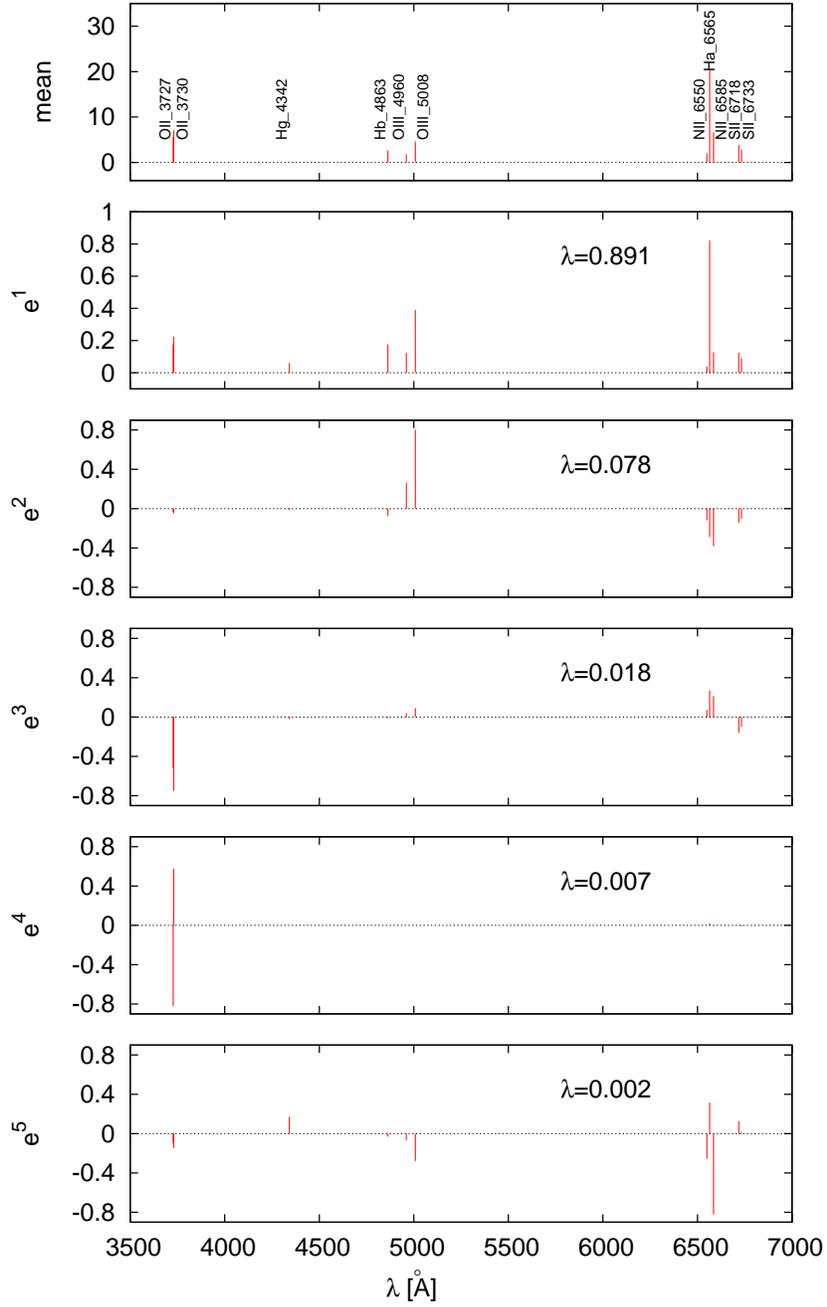}
\end{center}
\caption[]{Mean vector  {\ybar} and the first five eigenvectors of EW's. For each
  eigenvector, 
$\lambda$ 
denotes the relative information content. 
See explanation in Section~\ref{sec:eigenvectors}.}
\label{fig:ew_pca}
\end{figure}

Figure~\ref{fig:ew_pca} shows the results of PCA of the EW
 vectors.
The average EWs for each of the 11 lines 
(vector {\ybar}) are plotted in the top panel. Below these are the 
first five eigenvectors $\mathbf e^k$ ordered by their eigenvalues. Their 
information content is 89.1\%, 7.8\%, 1.8\%, 0.7\% and 0.2\%  of the total 
variance, respectively. The numerical results are summarized in Table~\ref{tab:eigv}.
The meaning of each PC can be easily interpreted by comparing the
weight of the $i$th line in the $k$th eigenvector $e_i^k$ with the
mean EW of the $i$th line $\overline y_i$.

The first eigenvector $\mathbf{e}^1$ is very similar to the average
vector $\overline \mathbf y$.  It means, the most important variation in
emission line EWs is simply a constant multiplicative
factor in the amplitude that varies from galaxy to galaxy. A larger $c_1$ indicates 
stronger nebular emission. The
PC is dominated by the strongest line {\Halpha}. 

The second eigenvector $\mathbf{e}^2$ represents mostly the {\OIIIdoublet} and
nitrogen lines. The coefficients of {\OIIIdoublet} 
and {\NIIdoublet} have opposite signs. The same holds for 
{\OIIdoublet} and {\NIIdoublet} in the eigenvector  $\mathbf{e}^3$.
This enables the nitrogen emission lines to vary independently
of oxygen in the reconstructed emission-line spectrum. The EW data
(Figure~\ref{fig:ew}) 
show
that [N\ionsign{II}] emission grows 
continuously, slowly from early to late types, while the oxygen lines have a
stronger type dependence becoming steep especially for the extremely
blue galaxies.
 Due to the higher ionization degree,  
the behavior of the 
 [O\ionsign{III}] doublet as a function of type is different from the
other lines: while showing no significant emission at low and moderate
{\eclass} values, there is a steep rise at {\eclass} $>0.5$. 
Nitrogen in these components is relatively strong compared to the
Balmer lines, thus both $e_2$ and $e_3$ influence the [N\ionsign{II}]/{\Halpha} ratio.

These eigenvectors do not  significantly change the ratios of lines in
the same 
doublet since their weights normalized by their average values are
nearly equal. The constant ratios have physical reasons and 
it is a strong effect which persists in these components.

In $\mathbf{e}^4$  the two lines of
 the {\OIIdoublet} doublet are represented with opposite signs. The
 effect of this component is to change their ratio. This eigenvector 
reflects the measurement errors of the \OIIdoublet, which is difficult
 to deblend with the resolution of SDSS spectroscopy.

Vector  $\mathbf{e}^5$  contains mostly
 \NIIdoublet, with some weak Balmer and oxygen lines. Thus we expect that it
 might influence the precise reconstruction of [N\ionsign{II}] lines. 

Doublet lines in some higher PCs often appear with opposite
sign, which is mainly the effect of errors.
We do not detail the further components as their variance is
 less than 0.2\% for each, they are dominated by noise.

\subsection{Eigenspace representation of emission line data}
\label{sec:eigenspace}

\begin{figure}
\begin{center}
\includegraphics{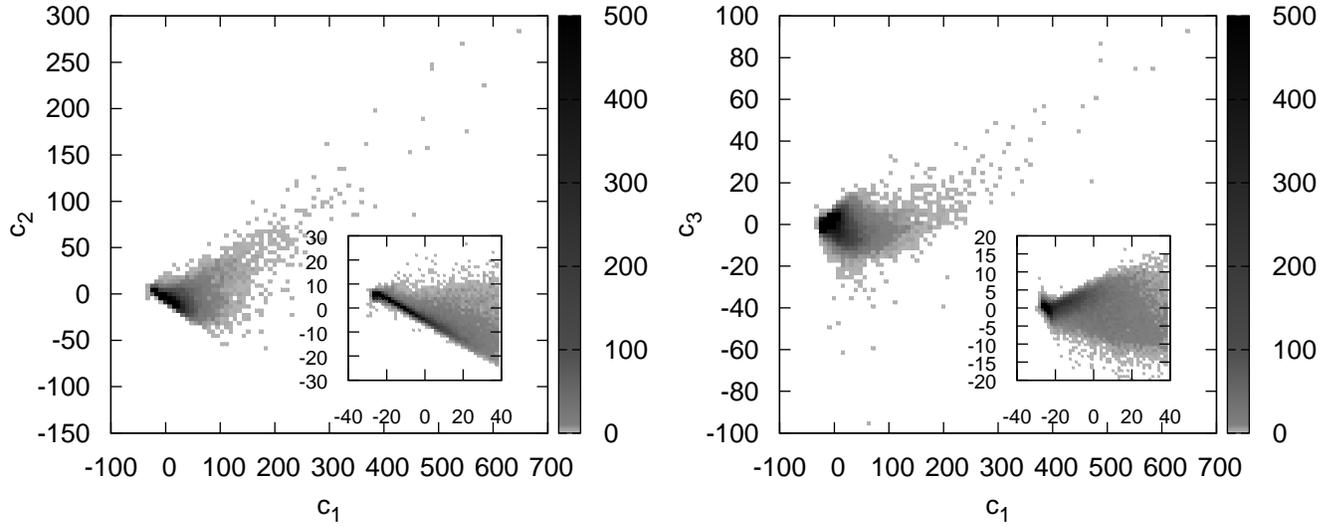}
\end{center}
\caption[]{Distribution of the emission line galaxies in the
  subspace of the first and second (left) and the first and the third (right) principal 
components. PCA shows that the data form a roughly 
two dimensional manifold  in the 11-dimensional EW space.
The inset plots show the low $c_1$ region zoomed in -  the
  distribution separates into a 'head' and a
  'tail'. }
\label{fig:pc-distr}
\end{figure}

The distribution of the data in the subspace of the first three
PCs is shown in Figure~\ref{fig:pc-distr}. 
The range occupied by galaxies is approximately a two-dimensional
curved manifold. 
Most of the objects form a triangular region that is
closely parallel to the axes  $\mathbf{e}^1$  and  $\mathbf{e}^2$ and
closely perpendicular to 
 $\mathbf{e}^3$, having small $c_3$ coordinates.  The distribution also has
a 'head' at the lower end in $c_1$ 
and a 'tail' 
having large $c_1$ 
values. The galaxies in the 'tail' also have significant contribution  from
the third PC but are still located on the curved surface
which is a continuation of the main locus described above.
The fact that PCA overestimates the dimensionality of the data is a 
known limitation of the method. It is because PCA is a linear transformation
 while the physics of emission lines generates
nonlinear structure (see Figure~\ref{fig:ew}).

\begin{figure}
\begin{center}
\includegraphics{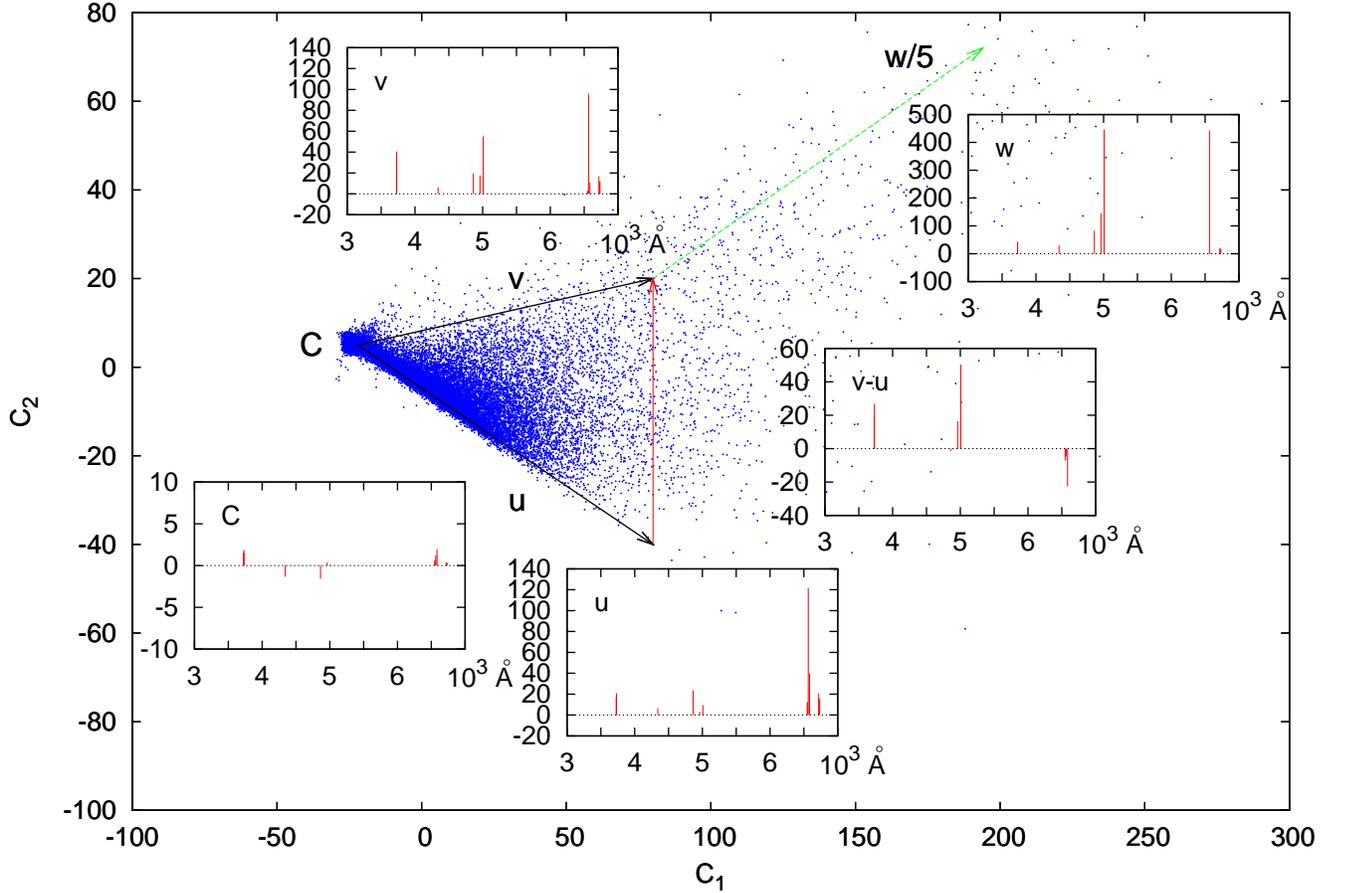}
\end{center}
\caption[]{EW data projected to the subspace of the first two principal
  components. Inset plots show 3D-reconstructed EW's corresponding to the origin
  $C$ and some representative directions. The two vectors 
$\mathbf{u}$ and $\mathbf{v}$ generate
  the main locus occupied by the majority of emission line
  galaxies. Their difference $\mathbf{v-u}$ contributes to the
  spectrum in the sense of enhancing oxygen, at the same
  time depressing nitrogen lines when going in the direction from
  $\mathbf{u}$ to  $\mathbf{v}$.
 The vector  $\mathbf{w}$  together with the previous two vectors
  generates the strongest emission  spectra. It 
 ends out of the range of this figure in the point $E$= (500,250,80),
the shown vector  $\mathbf{w}/5$ has the same direction and fifth the
  length of  $\mathbf{w}$.
The spectrum of point $E$ (not shown) is very similar to that of vector  $\mathbf{w}$.
Very strong Balmer lines and \OIIIdoublet, weak {\OIIdoublet}, as well as nitrogen
  deficiency can be observed.}
\label{fig:pca_vecs}
\end{figure}

As shown in Figure~\ref{fig:pca_vecs}, the points in the triangular
main locus embedded in the subspace of 
the first three eigenvectors can be generated by two vectors originated
in a point $C$ \citep{chan03}.
Their coordinates in ($c_1$, $c_2$, $c_3$) space are: $C=(-22,5,0)$,
$\mathbf{u}=(102,-45,15)$, $\mathbf{v}=(102,15,-15)$.
 The origin
and the vectors correspond to certain emission pattern recovered from
three PCs
 according to Equation~(\ref{eq:rec}).
As the figure shows, the origin has almost no emission. Thus, since the region is 
situated approximately in the ($\mathbf{e}^1$,$\mathbf{e}^2$) plane,
the EW values in these two vectors alone can give us some idea of
physical interpretation of the first two PCs.

\begin{figure}
\begin{center}
\includegraphics[width=17cm]{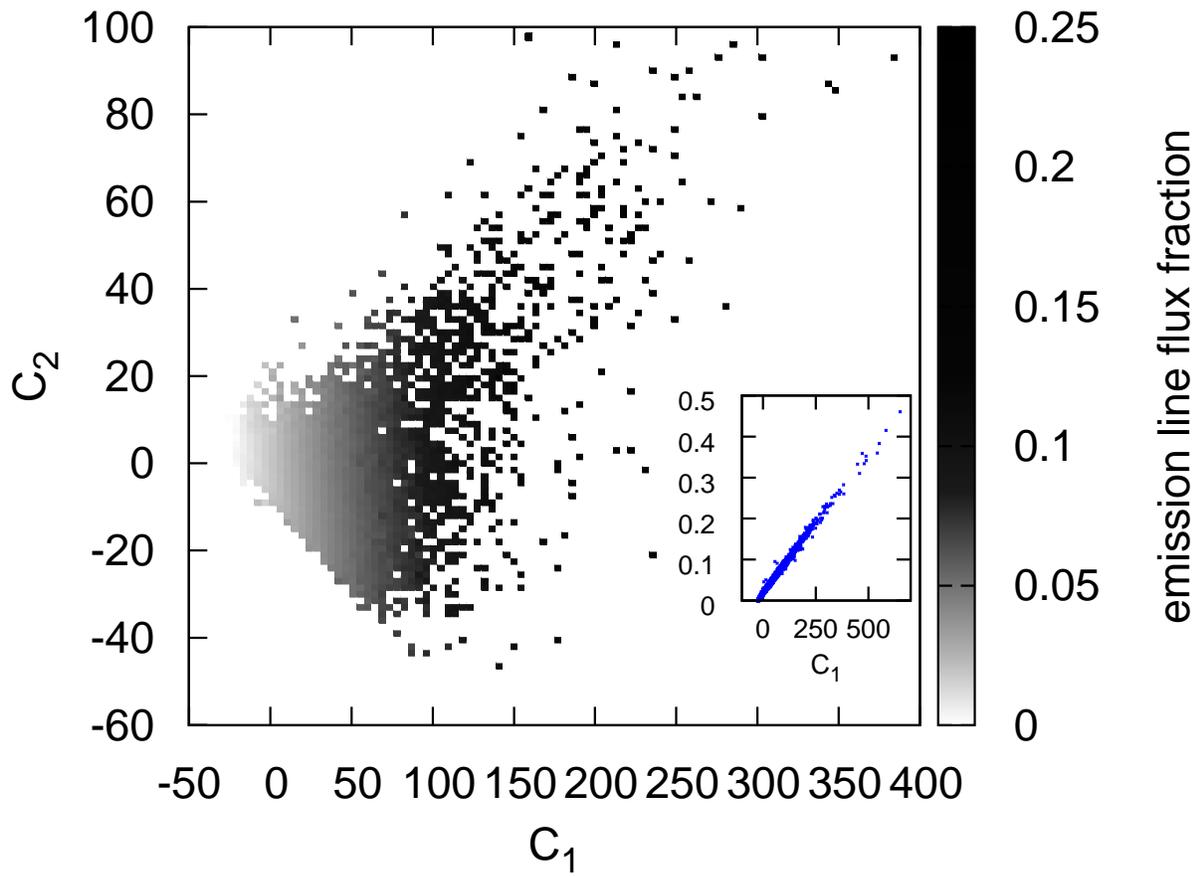}
\end{center}
\caption[]{Variation of the relative emission 
line flux $\mu$ 
averaged over bins (grayscale) shows 
a linear relation with $c_1$ (inset plot). 
\label{fig:pc-fluxratio}}
\end{figure}

As indicated in Section \ref{sec:eigenvectors}, the overall strength of the
emission is manifested in the first coefficient $c_1$. 
We define the relative emission line flux fraction $\mu$ as 
the total emission line flux normalized by the continuum flux within the range
$3728 - 6733$~{\AA}.
If we plot this quantity for each galaxy 
in the plane ($\mathbf{e}^1$,$\mathbf{e}^2$), 
we can see the monotonic growth of $\mu$ with $c_1$
(Figure~\ref{fig:pc-fluxratio}).
It is apparent 
that the gradient of $\mu$
is almost parallel to the axis $\mathbf{e}^1$. The inset plot shows
a linear relation between $\mu$ and $c_1$; the 'head' of the
distribution has an emission flux fraction of less than 1\%, while the
largest $c_1$ galaxies have $\mu\approx 0.5$, i.e., equal flux contribution
from continuum and 
emission lines in the analyzed wavelength interval.

\begin{figure}
\begin{center}
\includegraphics[width=17cm]{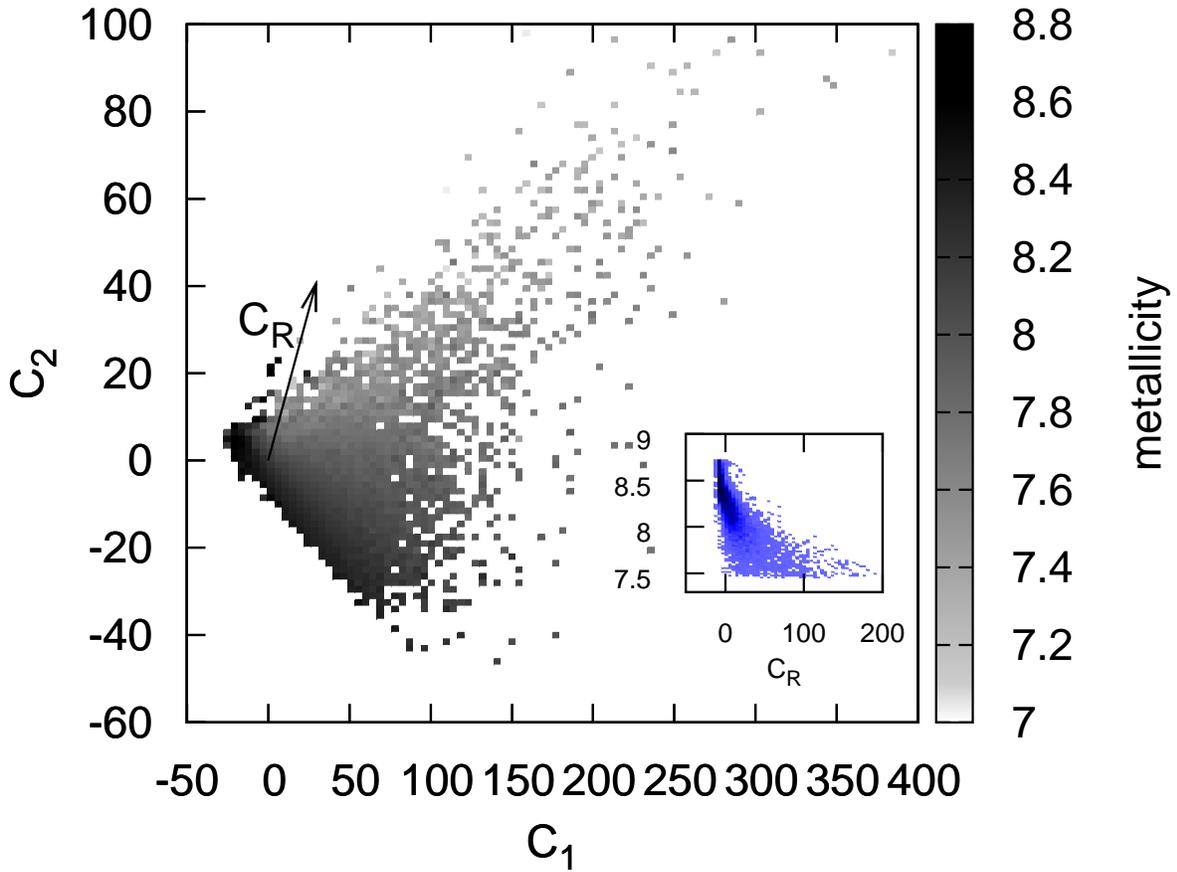}
\end{center}
\caption[]{Variation of metallicity with the first two PCA 
coefficients. 
The quantity 12+log(O/H)
estimated using Equation~(\ref{eq:metal}) 
averaged over bins is plotted in grayscale. The metallicity decreases in the 
direction of vector $\mathbf{C}_R$, which is close to vector $\mathbf{u}-\mathbf{v}$ 
of Figure~\ref{fig:pca_vecs}. The inset plot shows the distribution of metallicity along
$\mathbf{C}_R$.}
\label{fig:pc-metal}
\end{figure}

Another striking effect in Figure~\ref{fig:pca_vecs} 
is that while oxygen lines are relatively strong,
the nitrogen emission is 
weak in vector $\mathbf v$, $\mathbf {v-u}$, the difference, shows
negative {\NII} lines. This 
indicates a difference in metallicity. We estimated the ratio of oxygen and hydrogen
abundances from the ratio of {\NIIu} and {\Ha} fluxes using the
empirical formula of \citet{pettini04}.
\begin{equation}
12 + \log (\textrm{O/H}) = 8.9 + 0.57 \, \log \left(\textrm{\NIIu /\Ha}\right).
\label{eq:metal}
\end{equation}
As this estimator is calibrated for H\ionsign{II} galaxies we excluded
objects  dominated by non-thermal emission from this analysis by requiring
 $\log (\textrm{\NIIu /\Ha}) < -0.3$.
Figure~\ref{fig:pc-metal} 
shows that the vector $\mathbf {v-u}$ points approximately in the 
direction of the negative metallicity gradient:
at a fixed contribution from vector $\mathbf u$ the objects tend to
have smaller metallicities if the mixing ratio of $\mathbf v$ is
larger, i.e. upwards in the diagram. The 'head' has large
metallicity values which indicate old galaxies.

In Figure~\ref{fig:pca_vecs} data are plotted only up to
 $c_1=300$, however, the 'tail' 
 of the distribution reaches nearly $c_1=700$.
 The vector  $\mathbf{w}$  together with the previous two vectors
  generates the long tail. Its endpoint  $E$[\AA] =
  (500,250,80) is out of the range of the figure, so only a fifth of
 the vector, $\mathbf{w}/5$ is plotted. Point $E$ represents
  galaxies with extremely strong nebular emission.
Since vector  $\mathbf{w}$ carries very strong EW values compared to
 vector  $\mathbf{v}$ (or $\mathbf{u}$), the
  spectrum of the point $E$, as well as the points near the large
 $c_1$ end of the distribution, is very similar to that of vector
   $\mathbf{w}$. Very strong Balmer lines and \OIIIdoublet, weak
 \OIIdoublet, and nitrogen 
  deficiency can be observed. Even though our data are not  
corrected for reddening, the large  [O\ionsign{III}]/[O\ionsign{II}]
ratio is a real, strong effect. It implies a large  ionization
 parameter of the emitting gas, which increases  with $c_1$ at large $c_1$ values.
In summary, the galaxies of the 'tail'
have extremely
 strong emission, very low metallicities and high ionization parameters
 which indicates that  they are young bursting objects. 

\begin{figure}
\begin{center}
\includegraphics[width=12cm]{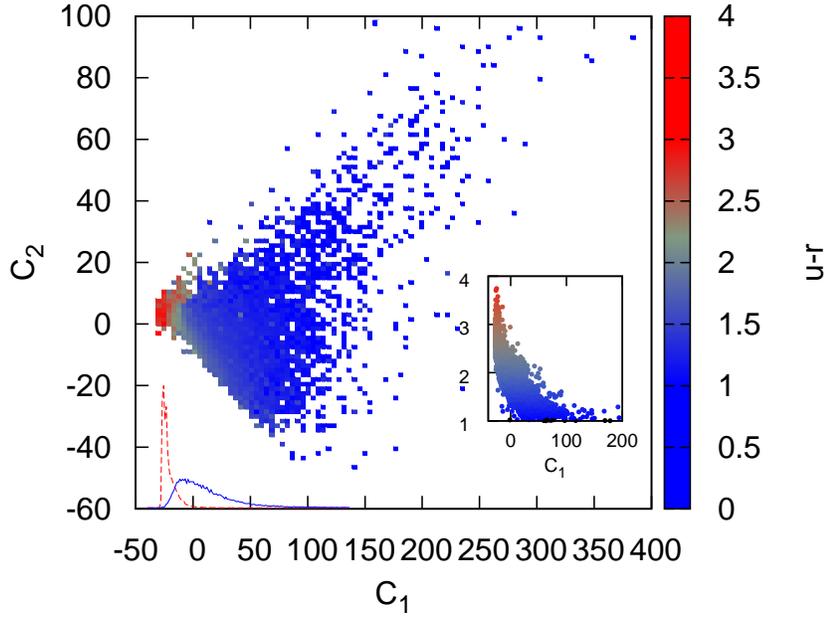}
\end{center}
\caption[]{Distribution of $u-r$ color in the plane 
($\mathbf{e}^1$,$\mathbf{e}^2$). Main plot: color averaged over
  pixels ($u-r$ color-coded, red means $u-r>2.22$, blue  $u-r<2.22$,
  gray is the transition between them). 
The lowest $c_1$ values are dominated by red objects, blue
  becomes dominant at higher $c_1$ values. Transition from red to blue 
  types ($u-r=2.22$) is around $c_1=-20$. Inset plot:  $u-r$ vs. $c_1$
  is a monotonic relation, the higher  $c_1$, the bluer
  objects. However, low $c_1$ ranges have mixed colors, the relation gets
  tighter at higher  $c_1$.
Histograms at the bottom: $c_1$ distribution of the two color
types. Red ($u-r>2.22$) subset plotted with red dashed line, blue
$u-r<2.22$ subset with blue solid line.
\label{fig:pc-color}}
\end{figure}

Figure
\ref{fig:pc-color} shows the color $u-r$ in the subspace of the first
and the second PC. A red 'head' and a blue 'tail' can be seen, and the color becomes continuously bluer toward large $c_1$ values. This indicates a
 close relation of the distribution in the emission line PC space to the
 bimodality seen in  color. The histograms 
 at the bottom of the diagram indicate the distribution of red
 and  blue galaxies defined by the  $u-r=2.22$ cut of
 \cite{strateva01}. The transition between the two distributions is
 around $c_1\approx -15$ where we have an equal number of red and blue
 galaxies in our sample. By a cut at  $c_1=-15$ we can select 
93\%/86\% of blue/red galaxies, with around 10\%/11\%
 contamination from the other  group, respectively.

\begin{figure}
\begin{center}
\includegraphics[width=12cm]{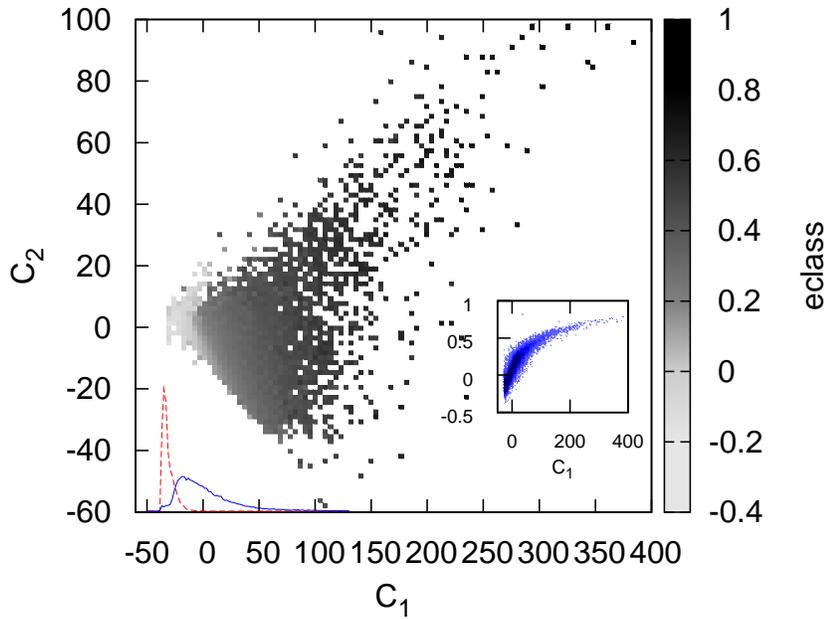}
\end{center}
\caption[]{Distribution of the spectral type  in the plane 
($\mathbf{e}^1$,$\mathbf{e}^2$). Main plot: {\eclass} averaged over
  pixels.  Inset plot:  {\eclass} vs. $c_1$. 
We can see the same tendency for {\eclass} as for color in Figure
  \ref{fig:pc-color}.
 Early types (negative \eclass{}) are at negative $c_1$ values, 
late types (positive \eclass{}) are at higher $c_1$.
Histograms at the bottom: $c_1$ distribution of the two spectral type bins. 
 Red dashed line: \eclass$<-0.05$, blue solid line: \eclass$>-0.05$.
\label{fig:pc-eclass}}
\end{figure}

\begin{figure}
\begin{center}
\includegraphics[width=15cm]{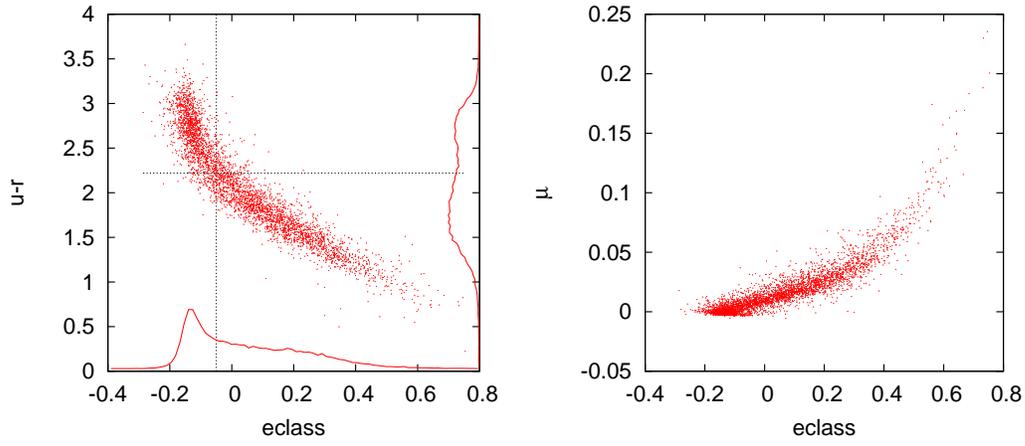}
\end{center}
\caption[]{Left panel: 
color $u-r$ vs. spectral type {\eclass}. The relation is nearly
linear, apparent bimodality. 
The $u-r=2.22$ separator of blue and red types (vertical dotted line)
  corresponds to \eclass{}$\approx=-0.05$ (horizontal dotted line). The
  histogram of the {\eclass} distribution is plotted on the $x$
  axis. The distribution of $u-r$ is projected to the right margin.
Right panel: connection between spectral type and relative emission
  parameter $\mu$. 
}
\label{fig:type-emi}
\end{figure}

The color--$c_1,c_2$ correlation indicates that $c_1$ must be strongly
correlated to the continuum  shape of the spectral energy distribution (SED). 
Figure~\ref{fig:pc-eclass}
shows the relation between {\eclass} and the first principal
components. The $c_1$--{\eclass} relation is very similar to the
relation of
$c_1$ and $-(u-r)$. This is
 not surprising as they both measure the same effect: the difference
between the intensity of the blue and the red end of the
spectrum. Their relation is illustrated in Figure
\ref{fig:type-emi} and resembles two linear relations, one for the
early type objects and the other for the late type objects. The $u-r=2.22$
cut in color corresponds to a cut 
at \eclass$\approx -0.05$. As shown by the histogram in
Figure~\ref{fig:type-emi} (bottom), it also roughly corresponds to the inflection 
point of the {\eclass} distribution. The color distribution is also shown
projected to the right margin of the diagram. In
fact, the separator lines lie somewhat blueward
from the inflection points in both {\eclass} and color, which might be
the effect of undersampling of early types by our selection. If
we consider \eclass$=-0.05$ as the separator of early  
and late spectral types, we can see that the separation
is even slightly clearer
than in the case of colors. The cut   selects 93\%/88\% of
late/early spectral types with a fraction of 8\%/11\% of early/late
type objects
misclassified by the cut.
The correlation of {\eclass} and the relative emission parameter $\mu$
is shown in the right panel of Figure~\ref{fig:type-emi}.
 We will discuss the issue of the  
 emission line PC's and spectral type   in more detail later in
 Section~\ref{sec:correl}. 

\begin{figure}
\begin{center}
\includegraphics[width=12cm]{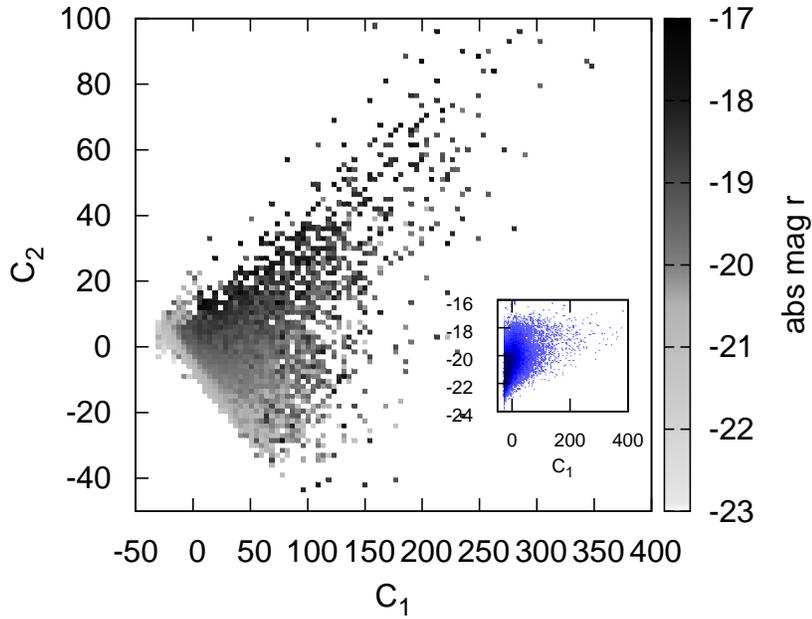}
\end{center}
\caption[]{$r$-band absolute in the plane of the first two PC's.
Main plot:  $M_r$ averaged over bins  
  (grayscale)  in the plane 
($\mathbf{e}^1$,$\mathbf{e}^2$). Inset plot: $M_r$ vs.
$c_1$. On average, luminosity decreases with increasing $c_1$,
however, the scatter is large.  
 The strongest nebular emission objects are
  the faintest ones, with $M_r \approx -17$.
\label{fig:pc-absmag}}
\end{figure}

The absolute magnitude in the $r$-band is shown in Figure
\ref{fig:pc-absmag}. Apart from the large scatter,  the 
objects are generally fainter at larger $c_1$. 
At the largest $c_1$ values only low
luminosity objects are present, with typical $M_r\approx -17$. This is
in concordance with the earlier studies  of SDSS galaxies by \cite{tremonti04} 
who found that the most metal deficient galaxies are faint.

\begin{figure}
\begin{center}
\includegraphics[width=15cm]{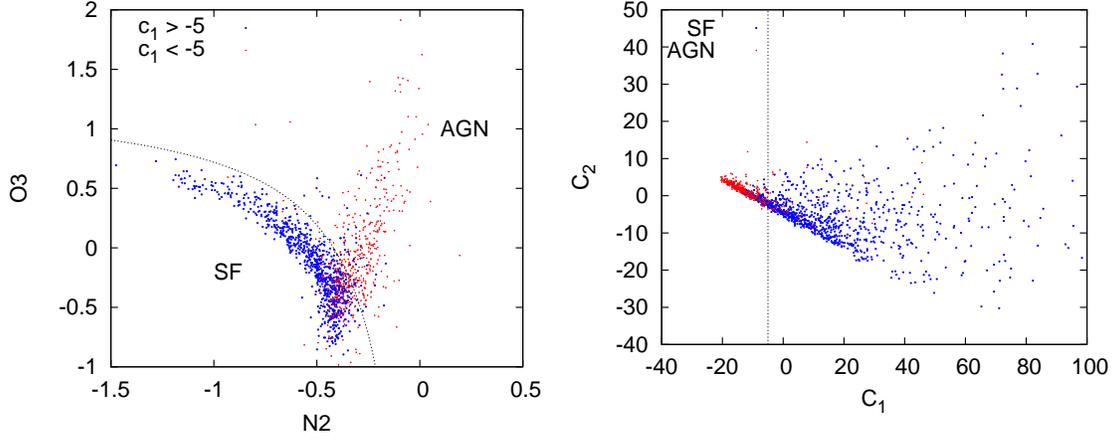}
\end{center}
\caption[]{Left:  N2:O3 diagnostic diagram for distinguishing between star
  forming galaxies and AGN. The dotted line shows the AGN separator
  of quation~(\ref{eq:agn-sep}). The two types of symbols are 
selected by $c_1=-5$ cut.
Righ: AGN (red) and SF (blue) defined by the separator n the
  right panel. AGN are situated at $c_1<0$. A part of the 'head' is
  missing because of some negative values among the EW's.
In both panels, the scatter plots show a 5\% random subsample.
}
\label{fig:seagull}
\end{figure}

Figure~\ref{fig:seagull} shows the connection of the first two PCs to
the AGN/SF diagnostic diagram. The line ratios 
$N2$ versus $O3$ are plotted in the
left panel, together with the 
AGN/SF separator line of Equation~(\ref{eq:agn-sep}).
The points are colored by the first principal component, using a  cut
at $c_1=-5$
which appears to be the most reliable $\mathbf c$-cut separating the  AGN from the SF. 
Red symbols denote $c_1<-5$, blue symbols indicate $c_1>-5$.  This
separator lies blueward of the color or spectral type separator, as a
significant fraction of AGNs has a bluer/later spectral  type and has
more contribution from emission lines than do the average red galaxies.
Up to the mixing at the lower edge of the
BPT diagram, the two $c_1$ regions roughly agree with the AGN/SF
separation: the $c_1$ cut selects 82\%/84\% of SFs/AGNs with 6.5\%/37\%
contamination from the other group, respectively.
We note that the 
$c_1=-5$ cut selects 92\%/58\% of SF/AGN as defined by \citet{stasinska06}, 
with
34\%/10\% contamination from the other group, respectively.
The cut of \citet{stasinska06} classifies more objects as AGN and is
therefore more consistent with a higher cut, about $c_1\approx 0$.
The separation of the types is not
well defined partly because of the mixing of SF and AGN activity in
some low emission galaxies.
In the right panel of Figure~\ref{fig:seagull} AGN (red) and SF galaxies
(blue), selected by Equation~(\ref{eq:agn-sep}) are plotted on the $c_1:c_2$
diagram. The vertical
 line is at $c_1=-5$. The plot shows the same subset of objects as in
 the left panel, which means all
galaxies  where Equation~(\ref{eq:agn-sep}) is not applicable are
excluded. They are low emission objects  having
non-positive \Ha, \Hb, {\NIIu} or {\OIIIu}. Nearly all  of
them  (99.9\%) are at $c_1<-5$. 
Note that low nebular emission galaxies are underrepresented
 due to the selection criteria. If present, the missing galaxies would
 populate the 
 'head' too. We also  note that AGN 
 with broad {\Ha} emission lines are 
 excluded from our sample. 
With our present sample, the 'head' (at $c_1<-5$) consists of low
emission objects, 31\% of which could be classified as AGNs. 
The plot confirms that the 'tail'
consists of SF galaxies. 
The classification using $c_1$ and $c_2$ is also reminiscent
of the (W(\Ha), \NII/\Ha) diagram 
proposed by \citet{fernandes10},
which distinguishes among SFs, AGNs and LINER-like galaxies, 
since $c_1$ is closely connected to the {\Ha} EW. However,
because of our selection criteria on equivalent width, we have 
very few LINER-like galaxies in our sample, these are situated at 
the lowest $c_1$ values.

Now we return to the question of missing low and high metallicity objects,
rejected because of the EW criterion.
According to a check carried out in both a low (7.5 -- 7.6) and a high 
(8.6 -- 8.8) metallicity bin, indeed, both ranges are underrepresented by about 
one-third in our selection, if compared to the rejected group.
However, there is still a sufficient number  of both low and high metallicity
galaxies in the selected sample for correlation analyses. 
Although for the relative occurence of the emission patterns our sample
might not be conclusive, it is still suitable for studying the relations
between emission lines and other observables.


We can conclude that  PCA isolates two groups of objects: red,
early spectral type, low emission, high metallicity, bright galaxies, with a
significant fraction of AGNs in the 'head' and the rest of the objects which are blue,
late spectral type, 
high emission, lower metallicity, fainter, star forming galaxies. In the
main locus, there is a gradient of all quantities listed
above. 
All these characteristics get continuously more prominent 
 and reach extreme values toward the end of the 'tail'.


The low effective dimensionality of the emission pattern found by PCA
 is not surprising.
As already discussed in the Introduction, the reason is the dominance of
only a couple of mechanisms. The largest impact on the emission line
flux pattern has the relative importance
of SF vs. AGN, the second most important influencing factor
is metallicity. These generate the effectively two-dimensional locus
of galaxies in the emission line space.

\subsection{Reconstructing spectral lines}
\label{sec:rec}

\begin{figure}
\begin{center}
\includegraphics[width=7cm]{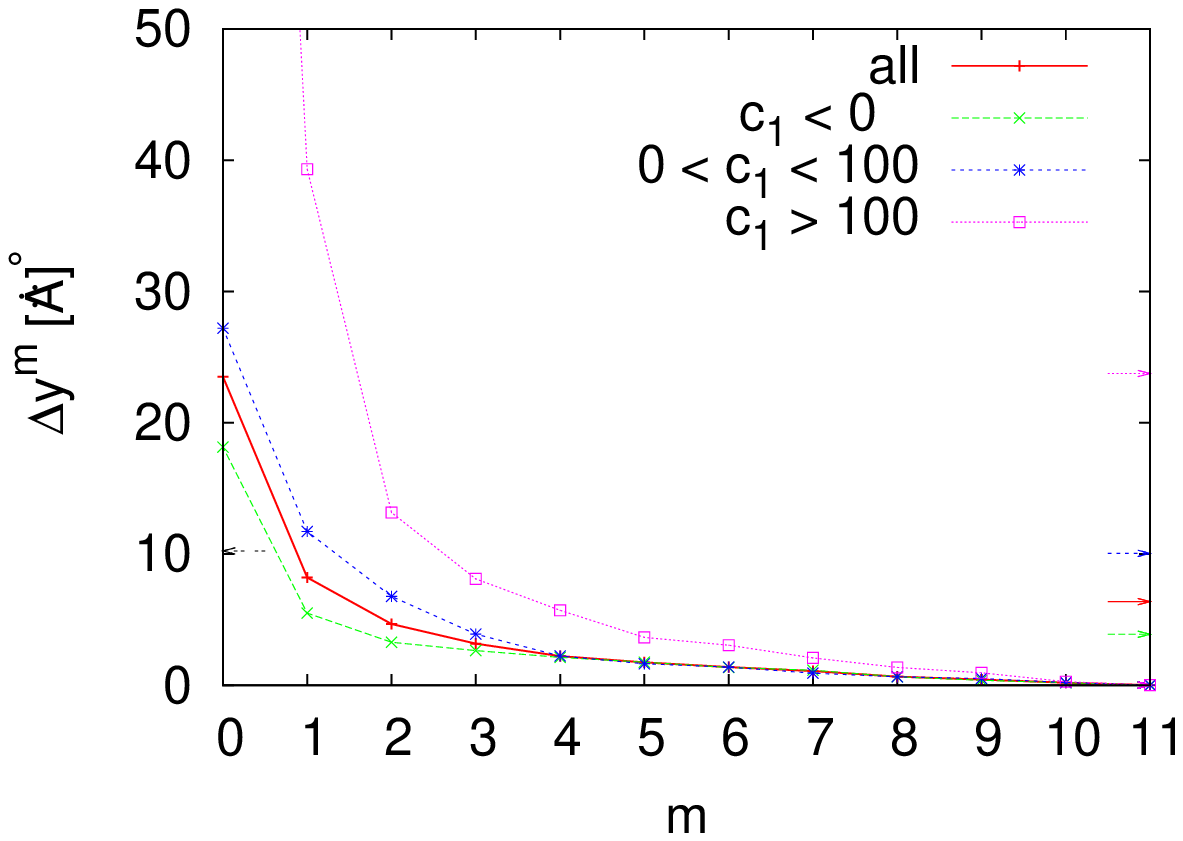}
\includegraphics[width=7cm]{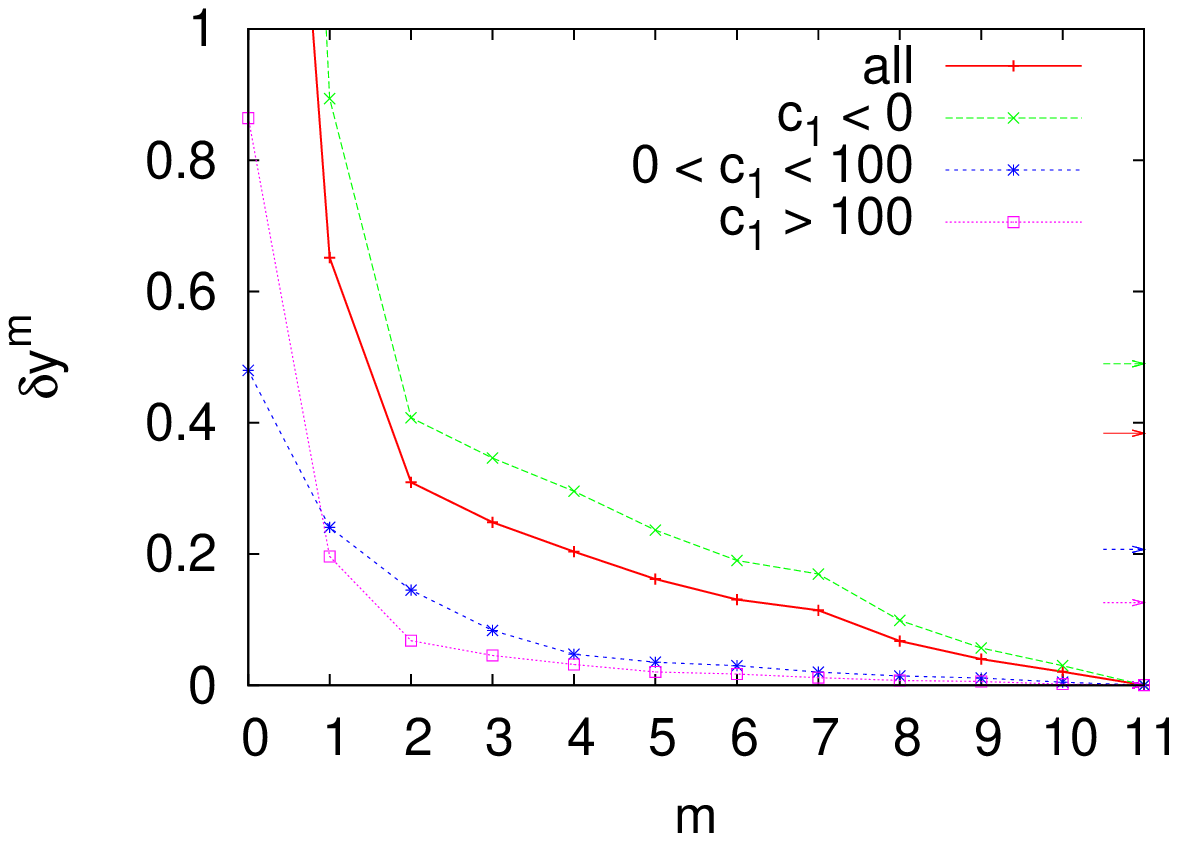}

\end{center}
\caption[]{Convergence of the spectral line reconstruction. The error of the reconstruction
as a function of truncation limit is plotted.
$m$: number of eigencomponents kept; $m=0$ represents the reconstruction using
the mean EW for each line, no PCs. Left: EW residuals
 summed over all lines. Right: relative error. The results
are shown for all galaxies (solid thick line) and for three $c_1$ bins; 
larger $c_1$ values indicate stronger nebular emission.
All EW's can be well reconstructed using the first three eigencomponents.
The single arrow at the left margin of the left panel denotes the
error of ignoring the emission line flux in \eclass$<-0.05$ early type objects,
see explanation in the text.
The arrows at the right margin show the same quantities for the
prediction made from continuum expansion coefficients, see explanation in Section
 \ref{sec:correl}.
\label{fig:chisqconv}}
\end{figure}

We studied the effect of truncation of the principal component basis on the 
restored EWs. 
We examined the convergence of the truncated EW estimator (Equation(\ref{eq:rec})) as
a function of the number of eigenvectors used for the reconstruction.
The error of the estimation can be characterized using the residuals
added in quadrature over all lines 
\begin{equation}
(\Delta  y^{(m)} )^2 = \sum_{i=1}^{11} ( y^{(m)} _i -  y _i)^2.
\label{eq:abserr}
\end{equation}
Figure~\ref{fig:chisqconv} shows $\Delta  y^{(m)} $ averaged over 
the whole sample as well as for three $c_1$ bins. 
Remember that larger $c_1$ values 
mean stronger emission and later spectral type. For the earliest bin
$c_1<0$ the contribution of the emission lines is so small that 
estimating by average ($m$=0 case) produces larger error than ignoring
the emission lines, i.e. setting EW=0 for all lines. We demonstrate the
error of the estimation by zero EWs for \eclass$<-0.05$ galaxies by
an arrow at the left margin of the diagram. For early spectral types,
ignoring the emission would mean a 10{\AA} error if summed over all lines.

If we drop all eigenvectors with eigenvalues  less than 1\%, 
we will have the first three PC's.
Their total
percent variance  for the case $m=3$ is 98.6\%. 
Using the first three components we can reconstruct
the total emission with 3{\AA} average precision. The error is type dependent, its
absolute value increases with $c_1$. For the strongest emission bin
the average $\Delta  y^{(3)}$ is 8{\AA}. 
However, unlike the absolute error, the relative
error defined as 
\begin{equation}
\delta  y^{(m)}  = \Delta  y^{(m)} / | y |
\label{eq:relerr}
\end{equation}
 is smaller for stronger emission objects. The average relative error
is 25\% for the whole sample, dominated by the error of  objects having
small EW values.
 For the extremely strong emission  bin
it is only 5\%. 

\begin{figure}
\begin{center}

\includegraphics[width=7cm]{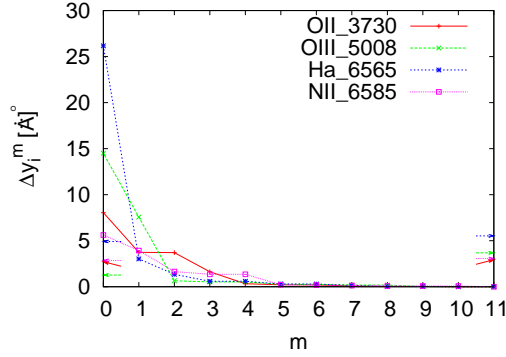}
\end{center}
\caption[]{Reconstruction of four selected emission lines using $m$
 eigencomponents. $\Delta y_i$: sample-averaged absolute EW 
error, $m$: same as in Figure~\ref{fig:chisqconv}.
The  arrows at the left margin show the error of estimation by zero
 emission line flux for the \eclass{}$<-0.05$ subset.
The arrows  at the right margin show the same quantity for the prediction made from
 continuum expansion coefficients, see explanation in Section \ref{sec:correl}.}
\label{fig:lineconv}
\end{figure}

We show the errors of the strongest,
 most important lines individually.
In Figure~\ref{fig:lineconv} we showed the convergence 
of the residuals of the individual EWs of  \OIIu, \OIIIu, {\NIIu} and \Halpha:
\begin{equation}
\Delta  y _i^{(m)} = |y^{(m)} _i -  y _i|
\label{eq:lineerr}
\end{equation}
averaged over the sample.
Similar to the overall emission represented by the summed residuals,
all the individual lines are 
reconstructed with an average error not larger than
2{\AA} using three PCs (or not larger than 4{\AA} using two PCs).
For the early type objects at \eclass$<-0.05$, we repeated the estimation by zero
emission line flux similar to Figure~\ref{fig:chisqconv}.
If we set all EWs to zero, the errors
 of the individual lines are still below 5\AA for this group, as shown by the errors
 in the left margin of the plot.

\begin{figure}
\begin{center}
\includegraphics[width=8cm]{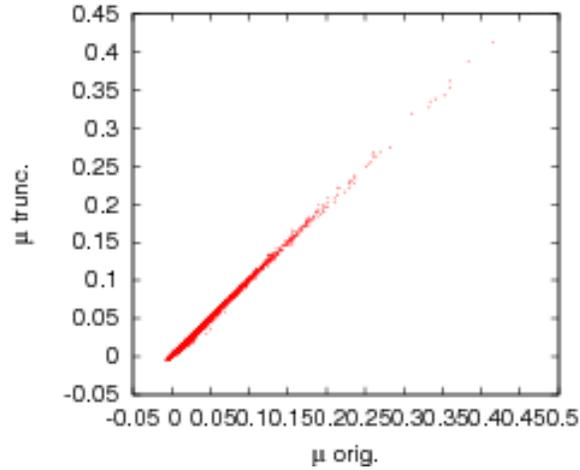}
\end{center}
\caption[]{
Reconstruction of the relative emission strength  using
truncated eigenbasis. The rms error is 0.001.}
\label{fig:trunc-fluxratio}
\end{figure}

\begin{figure}
\begin{center}
\includegraphics[width=8cm]{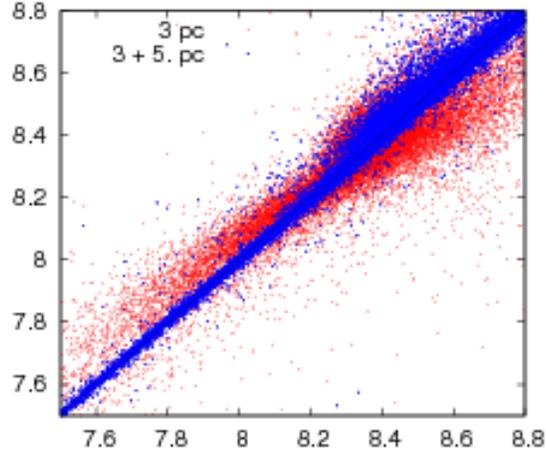}
\end{center}
\caption[]{
Reconstruction of metallicity using
truncated eigenbasis. With the first three eigenvectors, the error is relatively
large. Adding the fifth eigenvector suppresses the error
significantly.}
\label{fig:trunc-metal}
\end{figure}

We checked the effect of the truncation on physical quantities such as 
metallicity and emission flux fraction
if these were determined using three-PC-reconstructed EW data. 
The relative emission line flux fraction $\mu$ is plotted in Figure
\ref{fig:trunc-fluxratio}, truncated to the first three PCs versus the
original. The rms error of the reconstruction is as small as 0.001
which means an estimation of the emission line flux fraction within a precision
of 0.1\%. 
The lines reconstructed from the first three PC's obey the known ratios
of the doublet lines, especially {\OIIIdoublet} and {\NIIdoublet} 
with a relatively good precision. The line ratio {\NIIu}/{\NIIl} in the
reconstruction using the first three eigenvectors is 3.26, whereas the
fitted ratio in the 
original data is 3.23. For {\OIII}, the fitted  {\OIIIu}/{\OIIIl} ratio  is
 3.06 for the reconstructed data and 3.01 for the 
original data.
These features are so strong that only higher PC's begin
to violate them by including noise components. 
However, the flux ratios of lines that are not in the same doublet
are not restored  precisely. If we want to use 
the {\NIIu}/{\Ha} ratio for diagnostic purposes to distinguish 
between thermal emission of H\ionsign{II} regions from the AGN-like 
emission, we need the first three PCs and the fifth PC as
well. The fifth eigenvector is the one that makes 
possible an efficient fine-tuning of this ratio as it contains {\Ha} and
[N\ionsign{II}] with opposite signs.
Figure~\ref{fig:trunc-metal} shows that metallicity estimation with the
first three PC's has a relatively large error, which can be suppressed
by the inclusion of the fifth eigenvector. The rms errors of the
reconstructed metallicity are 0.23 and 0.13 without and with $\mathbf{e}_5$, respectively.
 The reconstruction is less precise at high
metallicities. This is because of the weak {\Ha} and {\Hb} lines, since,
as described in Section \ref{sec:eigenspace}, these are
typically early type galaxies.

\section{Correlation of spectral lines and continuum features}
\label{sec:correl}

The motivation of this study was  to explore the connection between the
continuum spectral type and the emission pattern of the emission line galaxies.
Although it is clear that there is no one-by-one relation, as a first step we
disregard the variations and focus on the systematic trends.  For the practical
applications, we would like to make predictions about theemission lines based on
continuum parameters. This can then be used, e.g., to add
emission lines to galaxy model SED's which only contain stellar
populations.

\begin{figure}
\begin{center}
\includegraphics[width=15cm]{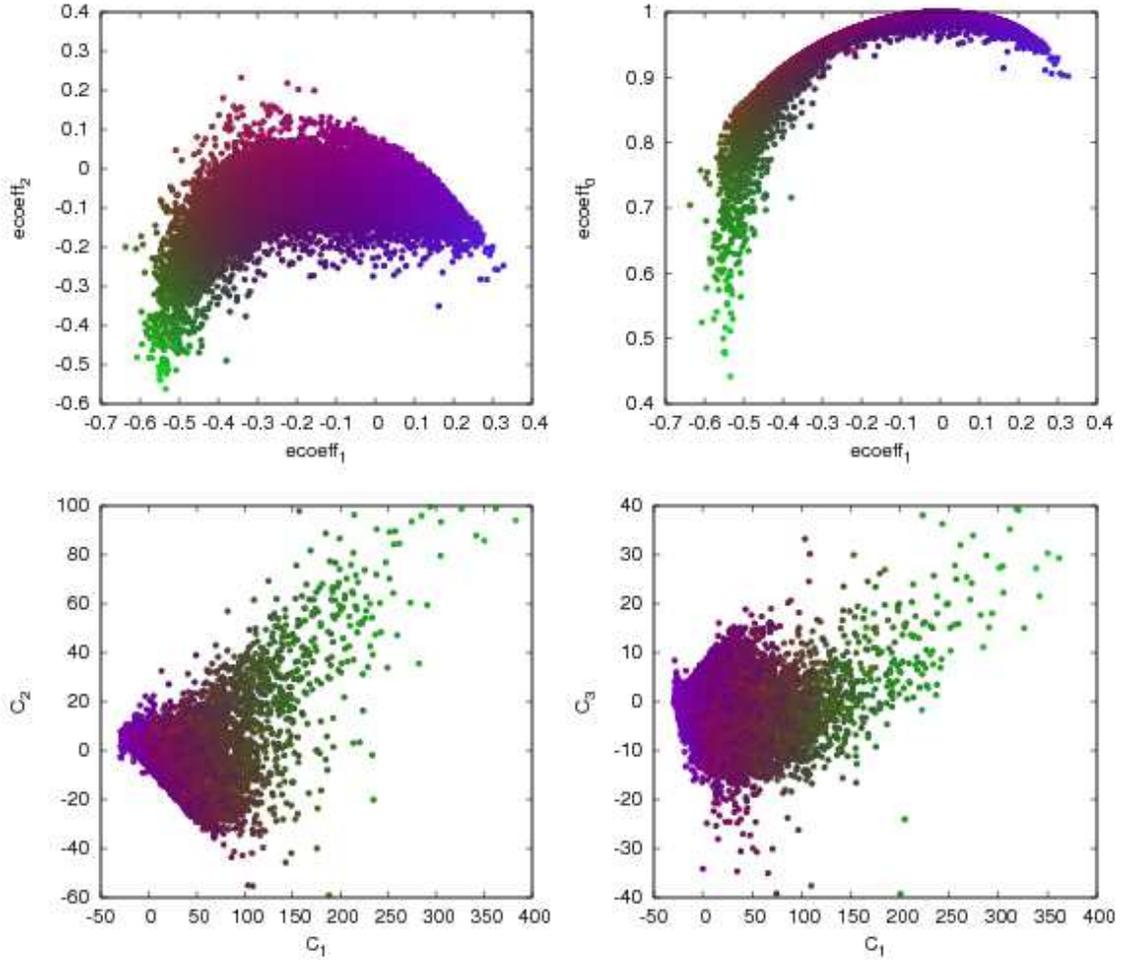}
\end{center}
\caption[]{Data points plotted in the planes \ecoeff{1}:\ecoeff{2}
  (top left), \ecoeff{1}:\ecoeff{3} (top right),
$c_1:c_2$ (bottom left) and $c_1:c_3$ (bottom left). The coloring is
  made by rgb-coding of \ecoeff{1} (green), \ecoeff{2} (blue),
  \ecoeff{3} (red). The same coloring is applied to the emission line PCA
  subspaces. By matching the points of the same colors in the various
  plots we can see how the \ecoeff{i} regions are mapped into the $c_i$
  space. The image shows, that (at least for \ecoeff{1}$<0.1$,
  which means positive \eclass, later type objects) 
there {\it is} a mapping. Early types are not resolved, however, they 
are located in the 'head' of the distribution, having weak emission lines.
\label{fig:ecoeff-pc-link}}
\end{figure}

We characterize the continuum spectrum with
the three most informative 
 coefficients of the spectral principal component expansion 
$\texttt{ecoeff}_0$, $\texttt{ecoeff}_1$ and $\texttt{ecoeff}_2$. We 
investigate their link to the first three 
emission line PCA coefficients $c_1$, $c_2$ and $c_3$ 
which proved to be essential in reconstructing
the emission with  sufficient accuracy.
Figure~\ref{fig:ecoeff-pc-link} illustrates how these parameters are 
linked to each other. The data points in all diagrams are colored by
\ecoeff{0}, \ecoeff{1} and \ecoeff{2}. For late type objects with
significant emission 
lines a mapping from the first three \ecoeff{}'s to the first three 
$c_i$ appears to be possible.

\begin{figure}
\begin{center}
\includegraphics[width=15cm]{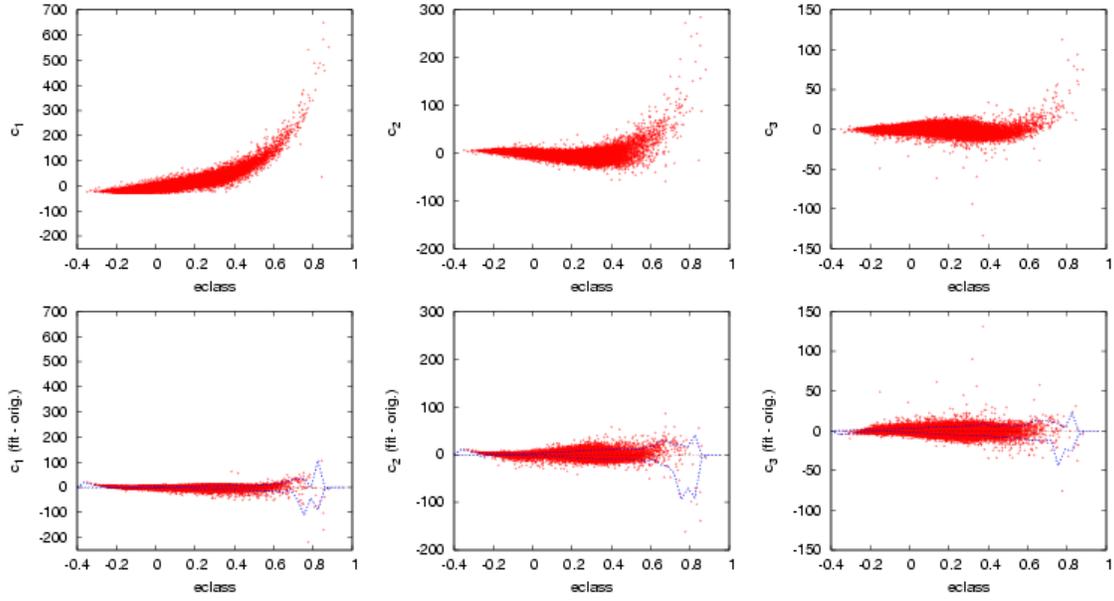}
\end{center}
\caption[]{Top: PC coefficients as a function of \eclass.
Bottom: the residuals of the first three emission line PC
  coefficients after subtracting the continuum fit, as a function of
  the continuum type $\texttt{ecoeff}$. 
An  empirical connection between the continuum 
spectral shape and the nebular emission pattern can be established from the 
observed correlations. The average residual scatter after subtracting the
  fitted estimator   is plotted with blue dotted lines
as  a function of the 
  spectral type.
\label{fig:ecoeff-pc-fit}}
\end{figure}

The correlation of each of the first three $c_i$ are plotted against
{\eclass} in the top panels of Figure
\ref{fig:ecoeff-pc-fit}. 
The first coefficient $c_1$ exhibits the strongest
correlation with the continuum features. This coefficient 
also has the largest information content. 
There are apparent systematic trends in $c_2$ and $c_3$ as well.
Given a continuum spectral type, we can determine the expectation
values and variances of the emission line EW's 
based on these empirical relations.
The ontinuum PCs \ecoeff{0}-\ecoeff{2} carry
even more information that can be used to establish an empirical relation.
 As indicated in the
previous section and as one can see in the top panels of 
Figure~\ref{fig:ecoeff-pc-fit}, 
early type galaxies might be fitted by constant 
values which would yield nearly zero flux. However, now we
choose to treat all 
data equally. We have checked that the two  approaches do not  
make a significant difference.
We fit a second order polynomial  of three variables
  $\texttt{ecoeff}_0$, $\texttt{ecoeff}_1$ and $\texttt{ecoeff}_2$
to each of $c_1$, $c_2$ and $c_3$,
\begin{equation}
c_i = \alpha_i + \sum_{k=0}^{2} \beta_i^{k} \texttt{ecoeff}_{k}
+\sum_{k=0}^{2}\sum_{l=k}^{2} \gamma_i^{kl}   \texttt{ecoeff}_{k} \,\texttt{ecoeff}_{l}.
\label{eq:fit}
\end{equation}
The fitted coefficients are listed in Table~\ref{tab:fitres}.
 We can
use this empirical relation to estimate emission properties solely
from continuum features of the spectra.
The  residuals of the fit, $c_i(\textrm{fit}) - c_i$, which characterize the
 goodness of the estimation,  are shown on the bottom panels of
Figure~\ref{fig:ecoeff-pc-fit}. 
 The rms error of the fit residuals is
plotted as a function of spectral type too.
 The origin of the scatter is mainly the
cosmic variance,
which includes
the effect of geometry (\cite{yip08}) and other
physical parameters not fully covered by 
the spectral classification parameter. The scatter becomes large toward
the largest {\eclass} values. However, as the flux values themselves
are large here, the resulting relative flux
error is smaller than for earlier types.

For practical applications, we also list the {\ecoeff{i}}
fits for four lines (\Ha, \OII, \OIII and \NII) in Table~\ref{tab:fitres_lines}.
For each doublet, we added up the values of the two individual lines.
However, this approach does
not recover flux information in such a compact form as
fitting the PCs. While with Table~\ref{tab:fitres} 
we restore
99.7\% of the emission line flux with just three components,
Table~\ref{tab:fitres_lines}  gives only  94.4\% of the flux with four fits
to a total of seven lines, and does not contain information on individual
lines except for \Ha.

We use the fitted $\tilde {c_i}$ values of Table~\ref{tab:fitres} 
to reconstruct the emission
lines analogously to Equation~(\ref{eq:rec}):
\begin{equation}
 \mathbf{\tilde{y}}
 = \sum_{k=1}^{3} \tilde{c}_k \, \mathbf{e}^k,
\label{eq:fitrec}
\end{equation}
We compare the emission lines coming from this estimator with the
measured values and calculate the errors described in Equations
(\ref{eq:abserr}) -- (\ref{eq:lineerr}), by substituting $\tilde y$
for  $y^{(m)}$.
The errors of this prediction are plotted
in Figures 
\ref{fig:chisqconv} and
\ref{fig:lineconv}
with small arrows at the right margin of each
plot. 
We find that the strength of the total nebular emission 
 can be predicted  from the spectral
continuum with an average accuracy of $5${\AA} or 40\% for the entire
sample. However, for the objects of greatest interest -- those
having significant emission -- the relative precision is better. The average errors
 are $10$\AA (20\%) for the $0<c_1<100$ bin, 
and $25$\AA ($\approx$10\%) for the strongest emission bin ($c_1>100$). 
Estimating by average only, without using the continuum dependence the
error can be as large as $\approx$~100\%.

\begin{figure}
\begin{center}
\includegraphics[width=15cm]{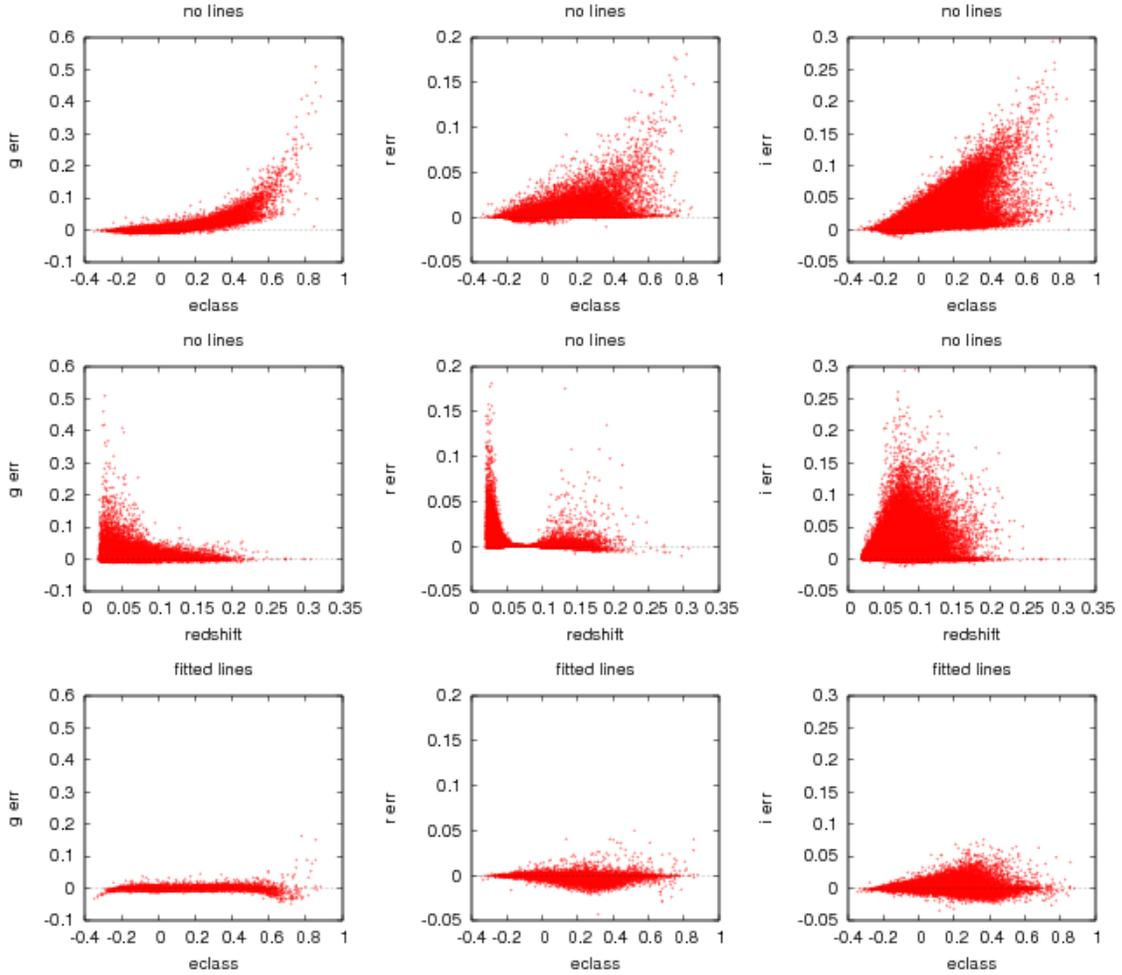}
\end{center}
\caption[]{
Difference in {\g}, {\r} and {\i} magnitude, if the photometry is made without
emission lines, as a function of spectral type (first row) and
redshift (second row). Late spectral types at larger {\eclass} having
stronger nebular emission will have larger errors. Redshift
dependence shows pattern caused by lines being redshifted into and out
from the filter. Third row: magnitude errors of the simulated photometry with emission
line from the continuum fit, as a function of spectral type.
\label{fig:photo-error}}
\end{figure}

We can estimate, how well the reconstruction of emission lines works
in  terms of photometry. We simulate photometry
by convolving the SDSS filters with the spectra of the objects. We
investigate the impact of emission lines by omitting them, convolving
just the continuum and comparing these magnitudes with the values obtained
from the entire spectrum (continuum+lines).
The results for the \g, \r, and {\i} bands are shown in the first and second rows of
Figure~\ref{fig:photo-error}. The impact of the nebular lines is
strongly type-dependent. For the strongest emission objects at
high {\eclass}, the magnitude difference due to the lack of emission lines
can reach $0.5^m$  in the {\g} band. This is the effect of
{\OIIIdoublet} and \OIIdoublet.  The largest difference  in the {\r} and 
{\i} bands is $\approx 0.2^m$.  
The significance of the emission line contribution becomes striking when compared
to the typical photometric uncertainties. (For an $r\approx 19^m$ galaxy the photometric
data have a typical error of $0.03^m$, $0.025^m$ and $0.07^m$ in the \g, \r and
{\i} bands, respectively.)
Note the redshift dependence due to lines
being redshifted into and out of the filters. Most apparent is the 
{\r} band, the low redshift hump comes mainly from {\Ha} 
and the high redshift hump from {\OIIIdoublet},
which are then
redshifted into the {\i} band. 
On the bottom three plots we estimated the emission lines from
\ecoeff{i}, using the fitted $\tilde c_i$ values and Equation~(\ref{eq:fitrec}). We
carried out simulated photometry with continuum + predicted lines, and compared this
with photometry simulated with real emission lines. The results show that the
prediction approximates the original values with a maximum error $\approx
0.1^m$ in
{\g} for the strongest emission objects and  $0.05^m$ for {\r} and {\i}.
The rms error is of the order of $0.01^m$ for the extremely high emission
bin \eclass$>0.6$ and  $\approx
0.001^m$  for \eclass$<0.6$. (We note that the early types at  \eclass$<-0.05$ 
have errors not larger than this even if no emission
line flux is added.)  This precision is sufficient for
the most photometric applications. For example, one can add spectral
lines to 
model SEDs of stellar populations or any empirical spectra with missing
emission lines in a way that makes the continuum
features and the emission pattern consistent with the observations.

As an example of practical applications a similar method has been successfully
applied to improve the spectral templates used in SDSS photometric redshift
estimation. In SDSS, we used a hybrid photometric redshift (photo-z) 
algorithm that holds the
advantages of both the template fitting and empirical approaches. Here we use
semi-empirical spectral templates that are based on the \citet{cww} (CWW)
empirical spectra with an additional extension of the UV and IR ends with
Bruzual-Charlot model spectra.  These are, actually, the most widely used
templates in the empirical template fitting photo-z applications.  In our
approach, the templates are iteratively trained on a reference set of galaxies
with known redshifts and photometry, so that they match better the photometry
of the SDSS data (\citet{budavari00}).  The quality of photo-z estimation
clearly improves by the use of repaired templates. However, we see
instabilities during the iterative procedure around the positions where our
paper's analysis would predict a strong line.  We attribute this effect to the
discrepancy between the continuum-line correlations of the CWW templates and
the SDSS data sample. To correct for this, we used a modified algorithm,
where the emission lines in the CWW spectra are replaced using the empirical
continuum-line correlations, discussed in this study.  This modified method was
indeed used in SDSS data releases DR4, DR5 and DR6 (\cite{dr4,dr5,dr6}), where
this ingredient has reduced the error on the photo-z estimate by 10\% for the
bluest galaxies.  In Table~\ref{tab:pz}, we show the effect of this procedure on
the photo-z results in detail. The photo-z type $t$ parameterizes the spectral
type, with 0 denoting the reddest galaxies, and 55 denoting the bluest galaxies. 
We list the rms photo-z error
as a function of type, for the cases when the procedure is carried out using
the original or the modified CWW spectra. While the estimate of the early type
objects remains unchanged as expected, there is an improvement increasing with
type.
 Beyond this, we consider the stabilization of the training
procedure as the main achievement.

We have shown how the relations shown in Figure~\ref{fig:ecoeff-pc-fit} can be used
to predict the expectation values of EWs based on the spectral type.
The variance may serve as additional information when a simulated distribution
of the emission pattern is generated.

\section{Conclusions} 
\label{sec:concl}

We considered a sample of over 40,000 SDSS galaxy spectra 
(emission lines and continua) coming mostly
from cores due to the aperture effect of the survey.
Using PCA of EWs of the 11
selected emission lines, we found that nearly 99\% of information is
included in the subspace generated by the first three
eigenvectors. They reconstruct emission line fluxes within a precision
of 5\%-25\%, depending on spectral type. We found that based on a
three-dimensional eigenspace representation of continuum spectra there
is a simple way of estimating the most probable emission line pattern
and its variation, which makes it possible to determine the total photometry 
from the continuum spectrum  in
the investigated bands with a precision $< 0.1^m$. 
The applications include the comparison of photometric
observations with models, e.g., determining K-corrections and absolute
magnitudes.
The prescription for adding lines on template spectra has
been successfully applied to improve the precision of photometric
redshift estimation.

\acknowledgments
The authors acknowledge
support from the following grants: OTKA-MB08A-80177,
MRTN-CT-2004-503929,  NKTH: RET14/2005, KCKHA005, and Pol\'anyi.

Funding for the creation and distribution of the SDSS Archive has been
provided by the Alfred P. Sloan Foundation, the Participating
Institutions, the National Aeronautics and Space Administration, the
National Science Foundation, the U.S.  Department of Energy, the
Japanese Monbukagakusho, and the Max Planck Society. The SDSS Web site
is http://www.sdss.org/.  The SDSS is managed by the Astrophysical
Research Consortium (ARC) for the Participating Institutions. The
Participating Institutions are The University of Chicago, Fermilab,
the Institute for Advanced Study, the Japan Participation Group, The
Johns Hopkins University, Los Alamos National Laboratory, the
Max-Planck-Institute for Astronomy (MPIA), the Max-Planck-Institute
for Astrophysics (MPA), New Mexico State University, Princeton
University, the United States Naval Observatory, and the University of
Washington.

\newpage


\clearpage

\newpage
\clearpage

\newpage

\begin{deluxetable}{ll}
\tablecolumns{2}
\tablewidth{0pc}
\tablecaption{\sc Analyzed Emission Lines}
\tablehead{
  \colhead{Line name} &
  \colhead{Rest wavelength [\AA]} 
}
\startdata
\OIIl   & 3727.09 \\
\OIIu   & 3729.88\\
\Hgamma & 4341.68 \\
\Hbeta  & 4862.68 \\
\OIIIl  & 4960.29 \\
\OIIIu  & 5008.24 \\
\NIIl   & 6549.86 \\
\Halpha & 6564.61 \\
\NIIu   & 6585.27 \\
\SIIl   & 6718.29 \\
\SIIu   & 6732.67
\enddata
\label{tab:linenames}
\tablecomments{Reference SDSS table, see also SpecLineName table in
the catalog science archive.}
\end{deluxetable}

\begin{deluxetable}{lrrrrr}
\tablecolumns{6}
\tablewidth{0pc}
\tablecaption{\sc The first five eigenvectors}
\tablehead{
  \colhead{Line name} &
 \colhead{\vtop{\hbox{\hspace{8mm}$e_1$}\vspace*{2mm}\hbox{\hspace{4mm}(0.891)}\vspace*{2mm}}}&
 \colhead{\vtop{\hbox{\hspace{8mm}$e_2$}\vspace*{2mm}\hbox{\hspace{4mm}(0.078)}\vspace*{2mm}}}&
 \colhead{\vtop{\hbox{\hspace{8mm}$e_3$}\vspace*{2mm}\hbox{\hspace{4mm}(0.018)}\vspace*{2mm}}}&
 \colhead{\vtop{\hbox{\hspace{8mm}$e_4$}\vspace*{2mm}\hbox{\hspace{4mm}(0.007)}\vspace*{2mm}}}&
 \colhead{\vtop{\hbox{\hspace{8mm}$e_5$}\vspace*{2mm}\hbox{\hspace{4mm}(0.002)}\vspace*{2mm}}}
}
\startdata
\OIIl   & 0.176 & -0.033 & -0.518 & -0.819 & -0.095  \\ 
\OIIu   & 0.225 & -0.046 & -0.750 & 0.574 & -0.148  \\ 
\Hgamma & 0.061 & -0.012 & -0.021 & 0.000 & 0.171  \\ 
\Hbeta  & 0.176 & -0.072 & -0.012 & 0.001 & -0.031  \\ 
\OIIIl  & 0.124 & 0.265 & 0.040 & 0.001 & -0.070  \\ 
\OIIIu  & 0.390 & 0.805 & 0.089 & 0.002 & -0.278  \\ 
\NIIl   & 0.039 & -0.118 & 0.072 & -0.001 & -0.254  \\ 
\Halpha & 0.821 & -0.289 & 0.271 & 0.018 & 0.315  \\ 
\NIIu   & 0.127 & -0.380 & 0.213 & -0.003 & -0.822  \\ 
\SIIl   & 0.125 & -0.143 & -0.159 & -0.004 & 0.127  \\ 
\SIIu   & 0.091 & -0.104 & -0.098 & -0.008 & 0.017  \\ 
\enddata
\label{tab:eigv}
\tablecomments{The first five EW principal components ordered by their relative
information content.
The eigenvalue of each eigenvector is given  in 
round brackets in the column header. }
\end{deluxetable}

\mycomment{
total percent variance:
0.891
0.969
0.986
0.993
0.995
}

\begin{deluxetable}{lrrrrr}
\tablecolumns{6}
\tablewidth{0pc}
\tablecaption{\sc The fit polynom coefficients of  $c_i$}
\tablehead{
  \colhead{\ } &
 \colhead{\vtop{\hbox{\hspace{8mm}$c_1$}}}&
 \colhead{\vtop{\hbox{\hspace{8mm}$c_2$}}}&
 \colhead{\vtop{\hbox{\hspace{8mm}$c_3$}}}&
}
\startdata
$\alpha$        & 1657.808 & 674.528 & 508.194  \\ 
$\beta_0$       & -4390.826 & -191.740 & -639.650  \\ 
$\beta_1$       & 488.087 & 108.530 & -27.376  \\ 
$\beta_2$       & -31.382 & -150.234 & -16.725  \\ 
$\gamma_{00}$     & 2717.923 & -481.908 & 132.141  \\ 
$\gamma_{01}$     & -610.956 & -59.598 & 25.392  \\ 
$\gamma_{02}$     & -153.612 & 226.695 & -2.644  \\ 
$\gamma_{11}$     & 546.841 & -498.071 & -245.997  \\ 
$\gamma_{12}$     & 212.847 & 82.486 & -36.546  \\ 
$\gamma_{22}$     & 433.339 & -617.698 & -375.677  \\ 
\enddata
\label{tab:fitres}
\tablecomments{Fitted polynom coefficients of equation~(\ref{eq:fit})
  for $c_1$, $c_2$ and $c_3$ as a function  of \ecoeff{i}.}
\end{deluxetable}

\begin{deluxetable}{lrrrrr}
\tablecolumns{6}
\tablewidth{0pc}
\tablecaption{\sc The fit polynom coefficients of  selected lines}
\tablehead{
  \colhead{\ } &
 \colhead{\vtop{\hbox{\hspace{8mm}\Ha}}}&
 \colhead{\vtop{\hbox{\hspace{8mm}\OIII}}}&
 \colhead{\vtop{\hbox{\hspace{8mm}\OII}}}&
 \colhead{\vtop{\hbox{\hspace{8mm}\NII}}}&
}
\startdata
$\alpha$	 & 1178.370503 & 2655.215341 & 308.461569 & -61.094836 \\
$\beta_0$	 & -3486.511250 & -4071.111890 & -1406.454947 & -443.963678 \\
$\beta_1$	 & 281.981021 & 1477.873084 & 625.325629 & -215.602859 \\
$\beta_2$	 & -100.191558 & -36.292149 & 38.422858 & -15.750437 \\
$\gamma_{00}$	 & 2315.964983 & 1415.882134 & 1103.062075 & 510.520196 \\
$\gamma_{01}$	 & -398.852773 & -1443.357079 & -664.083788 & 153.924612 \\
$\gamma_{02}$	 & -86.340819 & 40.335051 & -90.761428 & -77.564287 \\
$\gamma_{11}$	 & 550.149412 & -229.362689 & 579.736300 & 149.155350 \\
$\gamma_{12}$	 & 80.078273 & 309.940485 & 143.727637 & -85.912611 \\
$\gamma_{22}$	 & 454.181171 & -646.157142 & 663.971699 & 210.729024 \\
\enddata
\label{tab:fitres_lines}
\tablecomments{Fitted polynom coefficients of equation~(\ref{eq:fit})
  for {\Ha}, {\OIII}, {\OII} and {\NII} as a function  of \ecoeff{i}.}
\end{deluxetable}

\begin{deluxetable}{llll}
\tablecolumns{4}
\tablewidth{0pc}
\tablecaption{\sc The rms error of photo-z}
\tablehead{
  \colhead{} &
  \colhead{t} &
  \colhead{original} &
  \colhead{added lines}
}
\startdata
all & 0-55 & 0.0694 & 0.0680 \\
red & 0-20 & 0.0557  &  0.0556  \\
blue & 21-55 & 0.0876 & 0.0843  \\
bluest & 50-55 & 0.1512   & 0.1374 \\ 
\enddata
\label{tab:pz}
\tablecomments{
The rms error of photo-z for different galaxy types, for two different
templates sets. Column 3 lists the error of the redshift estimate when
the original CWW templates are trained and used as templates. Column 4
shows the results in the case when the emission lines are replaced 
before the training.
The type  parameter $t$ goes from 0 to 55, 0 being the reddest, 
55 the bluest type.}
\end{deluxetable}

\end{document}